\documentclass[onecolumn]{article}    % Enable this line and disable the 
\usepackage{graphicx}          % Include this line if your 
\usepackage{amssymb}
\usepackage{amsthm}
\usepackage{subfig}
\usepackage{mathtools}
\usepackage{authblk}
\usepackage{hyperref}

\usepackage{bm}
\usepackage{amsmath}
\usepackage{amsfonts}
\usepackage{subfig}

\usepackage{mathabx}
\usepackage{blindtext}

\usepackage{multirow}

\usepackage[a4paper, rmargin = 2.5cm, lmargin = 2.5cm]{geometry}

\renewcommand\vec{\mathbf}

\usepackage{float}
\usepackage{epstopdf}
\AppendGraphicsExtensions{.tif}

\title{Autonomous and Robust Orbit-keeping for Small Body Missions}

\author{Rodolfo Batista Negri \footnote{Ph.D. Candidate, rodolfo.negri@inpe.br, Av. dos Astronautas 1758, São José dos Campos.} and Antônio F. B. A. Prado \footnote{Head of Graduate Services, Av. dos Astronautas 1758, São José dos Campos. Volunteer Professor, Peoples' Friendship University of Russia, RUDN University. Associate Fellow AIAA.}}
\affil{National Institute for Space Research, 12227-010, São José dos Campos, São Paulo, Brazil}

\begin{document}

\maketitle

\section{Introduction}

%\lettrine{S}{ince} NEAR Shoemaker, an increasing number of missions aiming to explore small solar system bodies were performed. This interest is justified by many reasons, such as identifying the sources of water and organic molecules on Earth or learning more about our solar system's origins. Among these bodies, the asteroids - specially Near-Earth Asteroids (NEAs) - are of great interest because they are the main Earth impactors and a valuable source for in-situ resources near the Earth. However, the exploration of such bodies poses risks due to the low-gravity and highly perturbed environment, which, if not well approached, can lead to a dangerous path resulting in an undesired escape or impact \cite{antreasian2001design,scheeres2014close}.

The exploration of small bodies poses risks due to the low-gravity and highly perturbed environment, which, if not well approached, can lead to an unstable trajectory resulting in an undesired escape or impact. These conditions motivated many studies on guidance and control of spacecraft operating in their proximity \cite{guelman1994power,sawai2002control,broschart2005control,broschart2007boundedness,guelman2015closed,guelman2017closed,gui2017control,yang2017rapid}. In recent works, special attention is given to robust control laws (explicitly dealing with uncertainties), because of the perturbations, many of them using sliding-mode control theory  \cite{furfaro2013asteroid,furfaro2015hovering,yang2017finite}. The bulk of these works are focused on hovering approaches. However, some small body missions apply an orbital rather than a hovering approach as their main operation profile. For instance, the OSIRIS-REx mission, which is currently at asteroid Bennu, has two orbital phases in its close-proximity operation \cite{williams2018osiris}. An orbital profile, when well designed, allows for a stable long-term operation in close proximity to the small body, with little fuel expenditure. The orbit-keeping is made with small corrective maneuvers calculated by the ground team. An issue with the orbital mission profile is that the current stable orbits around small bodies are narrow due to the highly perturbed environment \cite{scheeres2013design}, placing limitations to the mission design and science outcome. Furthermore, the small body environment properties are not always known with sufficient accuracy before the spacecraft arrival, limiting the a priori design of stable orbits.

Although the ground-in-the-loop operation is flight-proven to be reliable, autonomous small body exploration is a rapidly advancing topic in astronautics. For instance, Takahashi \& Scheeres \cite{takahashi2021autonomous} show that an autonomous characterization of a near-earth asteroid (NEA) is feasible within current technology. Ohira et al. \cite{ohira2020autonomous} propose an optical navigation alternative that allows a spacecraft to navigate autonomously and in real-time around a small body. In this trend, an autonomous orbit-keeping control can reduce the mission operational cost and allow a more agile response to the environment, which in the ground-based approach can experience a delay of up to 20 minutes for NEAs. Indeed, when operating in proximity to distant small bodies, an autonomous operation might be the only solution, as the time delay in communication might make the operation unsafe if having the ground-in-the-loop. It also opens the possibility of stabilizing an orbit with active control, expanding the mission design possibilities and science outcomes.

Despite these advantages, few previous studies have focused on proposing autonomous orbital maintenance for small body missions. This is indeed justifiable because obtaining a reliable orbit-keeping law considering practical circumstances is not a straightforward task, as it should be: 1) robust, explicitly proven to deal with disturbances; 2) proved to be stable for any operation condition; 3) able to stabilize any orbital geometry; 4) preferentially able to control only the orbital geometry, so the spacecraft does not need to chase the two-body solution in time. The last condition means that the problem should be approached as a path-following control rather than reference tracking \cite{aguiar2007trajectory,aguiar2008performance}.

In a path-following approach, the spacecraft will only reconstruct the path (in this case, a Keplerian orbit) with no parameterization to be located in a specific point of the orbit at a predefined time. For the orbit-keeping problem, a path-following is more advantageous because it can easily accommodate periods in which the thrusters are turned off, allowing the orbit to vary under certain bounds before turning back on to bring the spacecraft to the reference orbit. In this process, controlling only the orbital geometry is the best approach as the spacecraft will not waste fuel chasing an unnecessary solution in time (i.e., the spacecraft will not try to match a specific true anomaly as the perturbations drift it from the reference orbit). It also avoids the disastrous outcome in which the spacecraft collides the small body when trying to match a position on the body's opposite side when the thrusters are turned on, with no need for an additional collision avoidance algorithm.

Guelman \cite{guelman2015closed,guelman2017closed} proposed a proportional-derivative feedback control that appears to be successful in attending the fourth point aforementioned because it only depends on geometrical parameters. Unfortunately, it is only suitable to circular orbits and not mathematically proven to be robust to disturbances. In recent work, Oguri \& McMahon \cite{oguri2021robust} achieve robustness in the operation by applying a stochastic control approach, but still formulate the problem as a reference tracking, and also relies on policies to be solved numerically, which adds complexity to the guidance. Other works not focused on small body missions \cite{schaub2000spacecraft,garulli2011autonomous,de2014virtual} have proposed the use of orbital elements and Gauss planetary equations for the orbit-keeping problem. However: the resulting laws are not robust; the stability cannot be proved when solving the resulting overdetermined system; the pseudo-inverse used in the control law is not proved to be non-singular (causing control peaks near the singularity); and it fails in attending condition 4, as the orbital element tied to the solution in time of the two-body problem is applied (the true anomaly, or an equivalent substitute).

In recent work, Negri $\&$ Prado \cite{negri2020path} derived and proved asymptotic convergence of a novel kind of path-following control. This control law is inspired by the solution of the two-body problem, where the eccentricity and angular momentum vectors are controlled to stabilize any Keplerian orbit. Although Negri $\&$ Prado \cite{negri2020path} derived and gave illustrative examples for different general applications, the fact that this path-following is inspired in the two-body problem makes it highly suitable for astronautics applications. Negri $\&$ Prado \cite{negri2020path} also showed that their path-following is robust to any bounded disturbances by using a unique set of sliding surfaces under the frame of sliding mode control theory. Moreover, the control law is entirely analytical and straightforward, being an excellent choice for operating in real-time.

This work applies an orbit-keeping control that satisfies all the aforementioned operational requisites for small body missions. They are met by using the robust path-following control derived by Negri $\&$ Prado \cite{negri2020path}. We show the advantages of such an approach for small body missions and how it could operate in practice, considering noises, modulation, idle-thrusters periods, and others. Our examples are mainly focused on the asteroid Bennu, which is currently visited by the OSIRIS-REx spacecraft, considering the actual noise level expected in the mission operation \cite{williams2018osiris}.

\section{Dynamical Environment}
\label{sec:dyn}

\begin{figure}
\centering
\includegraphics[width=.3\textwidth]{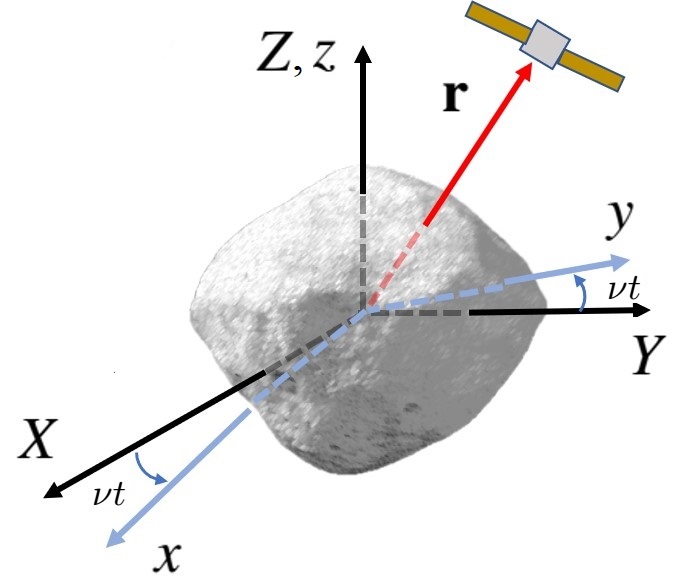}
  \caption{Inertial and body-fixed frames.}
 \label{fig:Iner_Body}
 \end{figure}

Close-proximity operations about small bodies can be simplified by representing the equations of motion in two different reference frames - inertial and body-fixed - according to the goal of the current task. Figure \ref{fig:Iner_Body} shows both frames centered in the small body center of mass. Capital letters are chosen to represent the inertial reference frame, and the lower-case ones are the body-fixed frame. For simplicity, we will assume that the spin axis of the small body is aligned with the inertial $Z$ axis. If both reference frames are coincident at the instant $t=0$, the rotation angle representing the misalignment between them is $\nu t$, in which $\nu$ is the spin velocity of the small body. The vector $\vec{r}$ will represent the position of the spacecraft in either of these frames. Therefore, the equations of motion are:

\begin{subequations}
\label{eq:System}
\begin{align}
\dot{\vec{r}}(t) &= \vec{v}(t), \\
\dot{\vec{v}}(t) &= \vec{f}(\vec{r},\vec{v}) + \vec{d}(\vec{r},\vec{v},t),
\end{align}
\end{subequations}
in which $\vec{v}(t)$ is the velocity vector, $\vec{f}(\vec{r},\vec{v})$ represents a known dynamics and $\vec{d}(\vec{r},\vec{v},t)$ is unknown disturbances acting on the system, but known to be bounded in magnitude by certain value.

In order to later stress our control law, we will assume that nothing but the main term of the gravitational field is known. In this case, the function $\vec{f}$ is simply $\vec{f} = -\frac{\mu}{r^3} \vec{r}$, with $\mu$ representing the gravitational parameter of the small body. The spin-state of the small body will be assumed to be unknown, so it is a disturbance that takes the following form in the body-fixed frame:
\begin{equation}
\vec{d}_{BF-\nu} = \begin{bmatrix}
2 \nu \dot{y} + \nu^2 x   \\
- 2 \nu \dot{x} + \nu^2 y  \\
0  \end{bmatrix},
\end{equation}

One of the main disturbances in a small body environment is the solar radiation pressure (SRP). Assuming that the sun is fixed in the $-X$ direction of the inertial reference frame, the solar radiation pressure in the body-fixed frame is, applying a cannonball model \cite{scheeres2016orbital}:
\begin{equation}
\label{eq:SRP}
\vec{d}_{BF-SRP} = \frac{ (1 + \rho)  P_0}{  B_{sc} S^2} \begin{bmatrix}
\cos(\nu t) \\
\sin(\nu t) \\
0
\end{bmatrix},
\end{equation}
with $\rho$ being the surface reflectivity, $B_{sc}$ the spacecraft mass to area ratio, the distance to the sun is $S$, and the constant $P_0=1\times 10^8$ (kg.km)/(s$^2$.m$^2$). 

Close to the asteroid, the gravity field's higher-order terms are the most significant perturbation acting on the spacecraft. In order to simulate these higher-order terms, we will employ a polyhedron gravity model to represent the gravity field of the small body, utilizing shape models from NASA's PDS Small Bodies Node website\footnote{\url{https://sbn.psi.edu/pds/shape-models/}}. Therefore, the disturbance acting in our system due to higher-order gravity terms is, in the inertial frame:
\begin{equation}
\vec{d}_{I-poly} = \begin{bmatrix}
\cos (\nu t) & - \sin (\nu t) & 0 \\
\sin(\nu t) & \cos(\nu t) & 0 \\
0 & 0 & 1
\end{bmatrix} \vec{g}_{poly} + \frac{\mu}{r^3}\vec{r}.
\end{equation}
The central body term is subtracted because it is already represented in $\vec{f}$. The force $\vec{g}_{poly}$ is obtained by using the method presented in Werner $\&$ Scheeres \cite{werner1996exterior}. We also make sure that the body-fixed frame is located on the small body's center of mass and aligned with the principal axes of inertia for the assumed mass \cite{dobrovolskis1996inertia}.

In some cases, prolonged simulations are needed, resulting in prohibitive computational costs if a polyhedron model is chosen to be applied. For such cases, we use a spherical harmonics model, with the well-known gravity potential: 

\begin{equation}
    U_{sh} = \frac{\mu}{r} \sum^{N}_{n=0} \sum^n_{m=0} \left( \frac{r_0}{r} \right)^n P_{nm}(\sin \varphi) \left[ C_{nm} \cos{(m\varrho)} + S_{nm} \sin{(m\varrho)} \right],
\end{equation}
where $\varphi$ and $\varrho$ are, respectively, the latitude and longitude of the spacecraft in the body-fixed frame, $P_{nm}$ are the associated Legendre polynomials, and $r_0$ is the reference radius. In order to keep coherence with the characteristics of the shape model, the coefficients $C_{nm}$ and $S_{nm}$ are calculated from the polyhedron shape \cite{werner1997spherical}. Recursive formulae are used to calculate the force from the gravity potential \cite{montebruck2000satellite}. For all the cases that we will later apply the spherical harmonics model, we calculate it up to the fifth degree and order.

\section{Orbit-keeping Control Law}

\subsection{Prolegomena}
\label{sec:prol}

The sliding mode control is a very intuitive concept. It lies in the fact that it is much easier to control a first-order system than a higher-order one~\cite{slotine1991applied}. The usual approach is to define a surface that is a linear combination of the states of the system~\cite{khalil2014nonlinear}. That is done to guarantee convergence to the desired state once the equilibrium condition for the sliding surface is reached \cite{utkin2017sliding}. In our case, the usual and straightforward approach would be to apply a reference tracking strategy, which, considering Eqs. [\ref{eq:System}], would use the conventional sliding surface: 
\begin{equation}
\label{eq:conv_slid}
\vec{s} = \tilde{\vec{v}} + \Lambda \tilde{\vec{r}},
\end{equation}
where $\Lambda>0$ is a diagonal matrix, and, $\tilde{\vec{r}}$ and  $\tilde{\vec{v}}$ represent the difference between the current position and velocity, respectively, to the desired ones ($\vec{r}_d$ and $\vec{v}_d$).

However, for the orbit-keeping problem, Eq. [\ref{eq:conv_slid}] presents the shortcoming of imposing a time parameterization for the movement on the orbit, which is often unnecessary. In our application, it would be more convenient to control only the orbit's geometry, which can be achieved with a path-following strategy. The path-following approach substitutes the parameterization in the time $t$ by a virtual arc-length $\theta$. 

\begin{figure}
\centering
\includegraphics[width=.9\textwidth]{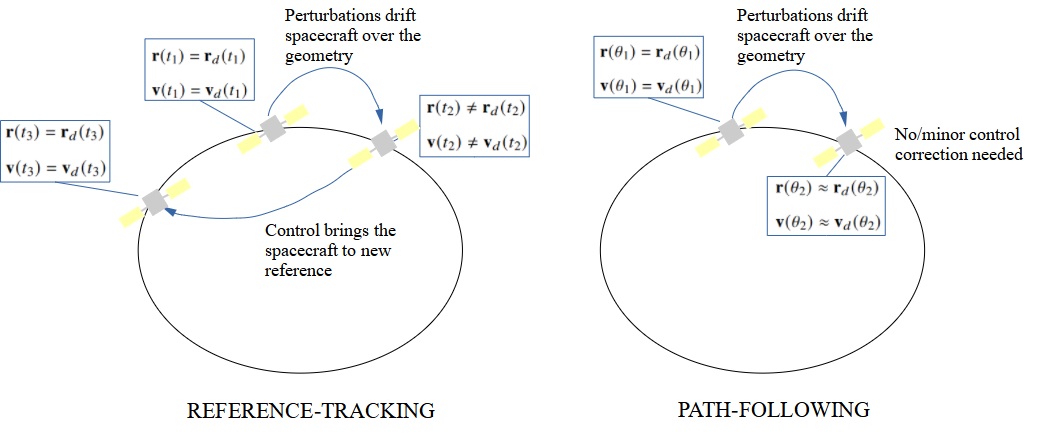}
  \caption{A schematic example of the advantage of the path-following control for orbital maintenance.}
 \label{fig:figurinha}
 \end{figure}

Let us exemplify some of the cases for which the path-following approach is more convenient than reference tracking. For instance, assume that a spacecraft is placed in an orbit that is stable in its geometrical orbital elements for a long period. For the sake of argument, let us assume that, for any reason, the spacecraft enters safe mode and loses its control commands for a short period. Now, assume that a hypothetical perturbation primarily disturbs the true anomaly, with the other orbital elements remaining close to the desired ones. When the autonomous spacecraft leaves safe mode, if no extra complexity is added to the reference tracking foreseeing such an event, the control will waste fuel trying to bring the spacecraft to the true anomaly of the corresponding time. On the other hand, the path-following would only compensate for the minor disturbances on the other orbital elements. Figure \ref{fig:figurinha} shows a schematic for that example.

Moreover, in autonomous orbital maintenance, it is required that the control has periods in which it is turned off, simply because the orbit might present good properties for an extended period, with no need for corrective maneuvers. It also provides time windows for the spacecraft to make scientific measurements without disturbances from the propulsive system. In other words, the spacecraft has to be allowed to diverge from the nominal orbit inside a chosen boundary before the thrust is turned back on. In that case, the path-following is also more straightforward than a reference tracking because it can easily operate in an on/off manner, with no extra complexity. A possibility when directly using reference tracking is in causing a disastrous outcome in which the spacecraft collides with the small body when trying to match a specific point of its orbit after the thrust is back on. For instance, let us assume that in a reference tracking strategy the two-body problem is simply being solved in time in order to obtain the reference signal for the control. It is clear that if the control is turned off, the perturbations might drift the spacecraft far from the reference. So, in this case, the control cannot be simply turned back on, as the reference state might be very far (e.g., on the other side of the small body). Therefore, if a reference tracking strategy is applied for the orbit-keeping, a second layer of complexity should be added to the algorithm to choose and update a new attainable reference. The control can be simply turned on in a path-following control because only the path is to be reconstructed, with no extra degree of complexity. Of course, we assume that the chosen path-following control is proved to operate within certain safe boundaries.

A first intuitive approach to tackle these issues, and to obtain a path-following law, would be to apply the Gauss planetary equations \cite{battin1999introduction}, defining the orbital elements tied to the orbital geometry as the spacecraft state and deriving a control law for them. However, five orbital elements define the geometry of the orbit, while the independent control variables are three. That means an overdetermined system, in which the most viable solution would be the least-squares, that guarantees no theoretical stability, as already noted by Schaub et al. \cite{schaub2000spacecraft} (using mean orbital elements and also the orbital element tied to the time solution). Furthermore, it is an issue guaranteeing that the matrix to take the pseudo-inverse is non-singular for all operation conditions, causing control peaks that may result in undesired behavior \cite{schaub2000spacecraft,garulli2011autonomous}.

\subsection{Radial-transverse-normal coordinates}
\label{sec:RTN}

Negri $\&$ Prado \cite{negri2020path} derived the control law using radial-transverse-normal (RTN) coordinates. The RTN is a right-handed coordinate system extensively applied in astrodynamics and celestial mechanics, in which the radial, transverse and normal components are defined according to the particle's kinematics. The radial component is in the direction of the position vector $\vec{r}$, the normal component is perpendicular to the orbital plane in the direction of the specific angular momentum $\vec{h}$, and, finally, the transverse component completes the right-handed frame. Their versors are written as a function of the particle's kinematics as:

\begin{subequations}
\label{eq:RTN_versors}
\begin{align}
\hat{r} &= \frac{\vec{r}}{r}, \\
\hat{\theta} &= \hat{h} \times \hat{r}, \\
\hat{h} &= \frac{\vec{h}}{h},
\end{align}
\end{subequations}
where $\vec{h}$ is the specific angular momentum vector and $h$ its magnitude. Therefore, an arbitrary vector $\vec{A}$ can be written in RTN as:

\begin{equation}
\label{eqn:write_in_RTN}
\vec{A}_{RTN} = \begin{bmatrix}
A_R \\
A_T \\
A_N
\end{bmatrix} = \begin{bmatrix}
\vec{A}\cdot \hat{r} \\
\vec{A} \cdot \hat{ \theta} \\
\vec{A} \cdot \hat{h}
\end{bmatrix}= \begin{bmatrix}
 \hat{r}^{\text{T}} \\
 \hat{ \theta}^{\text{T}} \\
 \hat{h}^{\text{T}}
\end{bmatrix} \vec{A},
\end{equation}
in which the superscript T represents the transpose, and the subscripts $R$, $T$ and $N$ stands for the radial, transverse and normal coordinates, respectively.

\subsection{Robust path-following Control for Keplerian Orbits}

For the reasons presented in Section \ref{sec:prol} the set of sliding surfaces defined in Negri $\&$ Prado \cite{negri2020path} is especially useful for the orbit-keeping problem. Those sliding surfaces only rely on the geometrical orbital elements, which have no time dependence in a two-body problem. The set of sliding surfaces using the definitions of RTN coordinates in Section \ref{sec:RTN} is:

\begin{equation}
\label{eqn:sliding_surface}
\vec{s} = \begin{bmatrix}
\tilde{\vec{e}} \cdot (\lambda_R \hat{r} + \hat{\theta}) \\
\tilde{h} \\
\hat{h}_d \cdot (\lambda_N \hat{r} + \hat{\theta})
\end{bmatrix}=0,
\end{equation}
where $\lambda_R>0$ and $\lambda_N>0$ determine the rate of convergence to the sliding surface, as shown in Proposition 1 in Negri $\&$ Prado \cite{negri2020path}, $\tilde{\vec{e}} = \vec{e}-\vec{e}_d$ represents the error between, respectively, the current and desired eccentricity vectors, $\hat{h}_d$ is the desired specific angular momentum versor and $\tilde{h}=h-h_d$ is the error in the magnitude of the specific angular momentum. 

The orbital plane can be unequivocally defined by the versor $\hat{h}$. Therefore, choosing an orbital plane means choosing a desired $\hat{h}_d$, or a desired inclination $i_d$ and longitude of the ascending node $\Omega_d$, as follows:

\begin{equation}
\label{eqn:ang_mom_ver}
\hat{h}_d = \begin{bmatrix}
\sin i_d \sin \Omega_d \\
- \sin i_d \cos \Omega_d \\
\cos i_d
\end{bmatrix}.
\end{equation}
The magnitude of the current and desired specific angular momenta can be respectively obtained by the expression:

\begin{equation}
\label{eq:angmom}
h = || \vec{r} \times \vec{v} || = \sqrt{\mu a (1-e^2)},
\end{equation}
in which $a$ is the semi-major axis and $e$ the eccentricity. Therefore, one can choose a desired semi-major axis $a_d$ and eccentricity $e_d$. Finally, the current and desired eccentricity vectors are respectively obtained by:
\begin{equation}
\label{eq:evec}
\vec{e} = \frac{1}{\mu} (\vec{v} \times \vec{h} - \mu \hat{r}) = e \begin{bmatrix}
\cos \Omega \cos \omega - \sin \Omega \sin\omega\cos i \\
\sin \Omega \cos \omega + \cos \Omega \sin\omega\cos i \\
\sin\omega \sin i 
\end{bmatrix},
\end{equation}
where $\omega$ represents the argument of periapsis, which implies that one can select the fifth and last geometrical orbital element $\omega_d$. 

Negri $\&$ Prado \cite{negri2020path} showed that, using the sliding surface in Eq. [\ref{eqn:sliding_surface}], robustness to bounded disturbances (i.e., each component of $\vec{d}$ in Eq. \ref{eq:System} is guaranteed to be bounded by certain value) and asymptotic convergence to the geometry of a Keplerian orbit can be obtained using the control:
\begin{equation}
\label{eq:control_theo2}
\vec{u}_{RTN}  = - F^{-1} ( G + K \text{sgn}(\vec{s})) - \vec{f}_{RTN},
\end{equation}
where $\vec{u}_{RTN}$ is the control command $\vec{u}$ in RTN, $K\in \mathbb{R}^{3\times 3}$ is a diagonal positive definite matrix, $\vec{f}_{RTN}$ is the known dynamics $\vec{f}$ written in RTN coordinates, the function $\text{sgn}(\vec{s}) \in \mathbb{R}^{3}$ represents the sign function taken in each component of $\vec{s}$, and the matrices $F$ and $G$ are defined by:
\begin{subequations}
\begin{align}
F &= \frac{1}{h\mu} \begin{bmatrix}
-h^2 & \left[2\lambda_R h-(\vec{v}\cdot\hat{r})r\right]h & -\mu r (\vec{e}_d\cdot\hat{h}) \\
0 & \mu rh & 0 \\
0 & 0& \mu r (\hat{h}_d\cdot\hat{h}) 
\end{bmatrix}, \\
G &=\frac{h}{r^2} \begin{bmatrix}
\tilde{\vec{e}}\cdot(\lambda_R\hat{\theta}-\hat{r}) -1 \\
0 \\
\hat{h}_d\cdot(\lambda_N\hat{\theta}-\hat{r})
\end{bmatrix}.
\end{align}
\end{subequations}

The control in Eq. [\ref{eq:control_theo2}] lies on the assumption that the magnitude of $\vec{h}$ and $\vec{r}$ are not zero. It also assumes that by defining an angle $\beta$ such that $\cos \beta = \hat{h} \cdot \hat{h}_d$, this angle is bounded by $\beta<90\degree$. As shown in Negri \& Prado \cite{negri2020path}, if both assumptions hold, the matrix $F$ is always invertible. Note that the violation of both assumptions would be highly uncommon in practice, especially in the orbital station-keeping problem. Nevertheless, if needed, both of them could be easily circumvented in an algorithm. The first, by making the spacecraft have any velocity not parallel to $\vec{r}$. The other, by choosing an intermediary $\hat{h}_d$ if $\beta\geq90\degree$.

Assume that the disturbances acting on the system are known to be bounded such that, for $\vec{d}$ written in RTN: $\lvert d_R \rvert < D_R$, $\lvert d_T \rvert < D_T$, and $\lvert d_N \rvert < D_N$. Thus, the diagonal gain matrix $K$ can be arbitrarily chosen, guaranteeing convergence, as long it satisfies the following inequality \cite{negri2020path}:

\begin{subequations}
\label{eqn:K_matrix}
\begin{align}
K_{1,1} &\geq  \frac{h}{\mu} D_R + \left\lvert \frac{2\lambda_R h-(\vec{v}\cdot\hat{r})r}{\mu} \right\rvert D_T + \frac{ r \left\lvert \vec{e}_d\cdot\hat{h}\right\rvert}{h}  D_N, \\
K_{2,2} &\geq r D_T, \\
K_{3,3} &\geq r \frac{\hat{h}_d\cdot\hat{h}}{h} D_N,
\end{align}
\end{subequations}
in which $K_{j,j}$, $j=1,2,3$, are the diagonal elements of the matrix $K$. Finally, the control $\vec{u}$ in the frame where $\vec{r}$ and $\vec{v}$ are defined is simply calculated as:
\begin{equation}
    \vec{u} = u_R \hat{r}  + u_T \hat{\theta}  + u_N \hat{h}  .
\end{equation}

\subsection{Practical Considerations}
\label{sec:pract}

It is widely known that the main drawback of the sliding mode control is its discontinuous control input, which in many practical applications leads to chattering. The discontinuity can be avoided by allowing the system to converge to a boundary around the sliding surface, at the expense of a bit of performance, as extensively documented in the literature \cite{slotine1991applied,utkin2017sliding}. The most common approach is to substitute the sign function with the saturation function:
\begin{equation}
\label{eq:sat}
\text{sat}(x;x^\star) = \begin{cases} 1, &\text{$x>x^\star$} \\\frac{x}{x^{\star}}, &\text{$-x^{\star}\leq x\leq x^\star$} \\ -1, &\text{$x<-x^\star$} \end{cases}.
\end{equation}
In the case $\vec{x}$, $\vec{x}^\star \in \mathbb{R}^n$, we define $\vec{g}=\text{sat}(\vec{x};\vec{x}^\star): \mathbb{R}^n \xrightarrow{} \mathbb{R}^n$, such that $g_i=\text{sat}(x_i;x_i^\star)$, i = 1,2,...$n$, for $i$ representing each component of the vector. Therefore, Eq. [\ref{eq:control_theo2}] can be restated as:

\begin{equation}
\label{eq:u_sat}
\vec{u}_{RTN}  = - F^{-1} \left( G + K \text{sat}\left(\vec{s}; \vec{\Phi}\right)\right) -\vec{f}_{RTN} ,
\end{equation}
in which $\Phi\in \mathbb{R}^3$ is a design parameter chosen by the designer. Specific values of $\Phi$ are defined later in Section \ref{sec:analysis}.

\begin{figure}
\centering
\includegraphics[width=.3\textwidth]{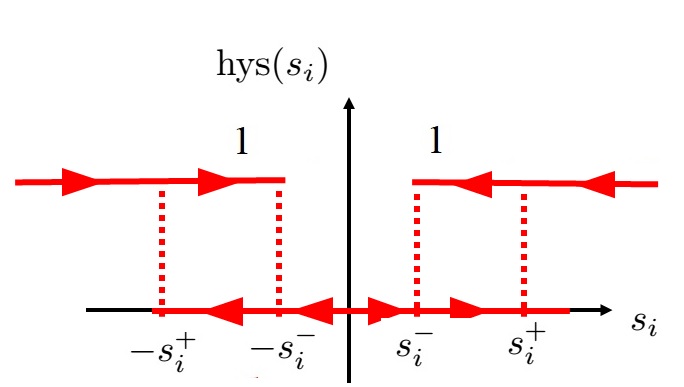}
  \caption{The hysteresis function for each component of $\vec{s}$.}
 \label{fig:hys}
 \end{figure}

As discussed in Section \ref{sec:prol}, it would be desirable that the control law accommodate turned off periods. This could be accomplished with different approaches, depending on the mission goals and available hardware. For illustrative purposes, we choose to apply a hysteresis function inspired in the Schmitt trigger, defined as:
\begin{equation}
\label{eq:hys}
\text{hys}(x) = \begin{cases} 1, & \text{$|x|>x^+$}  \\
0, & \text{$|x|<x^-$} \\
1, & \text{$x^-\leq |x| \leq x^+$ and hys($x_{p}$)$=1$}  \\
0, & \text{$x^-\leq |x| \leq x^+$ and hys($x_{p}$)$=0$}  
 \end{cases},
\end{equation}
where $\text{hys}(x_p)$ indicates the previous value of the hysteresis function. For $\vec{x} \in \mathbb{R}^n$ we define $\vec{g}=\text{hys}(\vec{x} ): \mathbb{R}^n \xrightarrow{} \mathbb{R}^n$, such that $g_i=\text{hys}(x_i)$, i = 1,2,...$n$. Therefore, a practical form for the path-following control law, considering idle-thrusters periods, can be:

\begin{equation}
\label{eq:u_sat_hys}
\vec{u}_{RTN}  = - \left[ F^{-1} \left( G + K \text{sat}\left(\vec{s}; \vec{\Phi}\right)\right) - \vec{f}_{RTN} \right] \lvert \lvert  \text{hys}\left(\vec{s}\right) \rvert \rvert_\infty,
\end{equation}
where $\text{hys}\left(\vec{s}\right)$ represents the hysteresis function taken in each component of the vector $\vec{s}$, as shown in Fig. \ref{fig:hys}. The operator $\lvert \lvert \cdot \rvert \rvert_\infty$ is simply the $\mathcal{L}_\infty$ norm, i.e.  $\lvert \lvert \vec{x} \rvert \rvert_\infty$ = $\max_i \lvert x_i \rvert$. This is just an ingenious mathematical way of saying that the control is turned off if all elements of $\text{hys}\left(\vec{s}\right)$ are zero, and turned on if any component of $\text{hys}\left(\vec{s}\right)$ is 1.

We should note that the sliding surface $\vec{s}$ can be written in terms of orbital elements, providing the designer a way of translating acceptable bounds in the orbital elements to the hysteresis in $\vec{s}$. That can be done by using Eqs. \ref{eq:RTN_versors}, \ref{eqn:ang_mom_ver}, \ref{eq:angmom}, and \ref{eq:evec} with the fact that $\hat{r} = \cos \alpha \hat{e} + \sin \alpha \hat{e}_\perp $, where $\alpha$ represents the true anomaly, $\hat{e}=\vec{e}/e $, and $\hat{e}_\perp = \hat{h} \times \hat{e}$. Although the transformation does not allow to obtain an explicit transformation from the bounds of orbital elements to $\vec{s}$, mainly because of the angular elements, it is still a straightforward procedure to be made with graphical tools within the mission orbital envelope. Furthermore, Negri and Prado \cite{negri2020path} show that, from the bounds in $\vec{s}$, it is guaranteed that the spacecraft will operate within certain bounds in $r$, which can be useful to guarantee that no impact will occur.

% colocar na frase althougt the transf etccccc \footnote{Because it is a pretty straightforward procedure, and tremendously paper-length consuming if made here, we consider that there is no scientific interest in presenting an example.}

\section{Analysis and Discussion}
\label{sec:analysis}

This section aims to show some illustrative examples to present how this path-following control can be applied for small body missions and its advantages. In all of the examples presented here, we will calculate the gain matrix $K$ from the equality satisfying Eqs. [\ref{eqn:K_matrix}], for the assumed disturbance bound $D=D_R=D_T=D_N$, and the constants $\lambda = \lambda_R=\lambda_N$. The vector $\vec{\Phi}$ is always calculated as $n_{\Phi}$ times the correspondent value of the diagonal of the matrix $K$, i.e. $\vec{\Phi}= n_\Phi \text{diag}(K)$. We also set the control update rate as 0.25 Hz (i.e., every 4 seconds), unless noted otherwise for specific cases. One certainly notes that this control update frequency is unnecessarily low, as the current spacecraft could calculate and execute the control in a larger frequency. That is more due to our computational limitations to run the simulations, especially because the polyhedron model is computationally demanding. Nevertheless, this is not detrimental to the presented applications, as we will show soon. This downgrade highlights the precision and accuracy of the control law. As we show later, the results will be more precise, and a much smaller boundary around the sliding surface can be chosen for the saturation function defined in Eq. [\ref{eq:u_sat}], with a smaller $\vec{\Phi}$ \cite{slotine1991applied}, if a larger control update frequency is applied. We inform the reader that a video for each of the simulations is uploaded in the main author's web page \footnote{They can be accessed in: \url{rodolfobnegri.com/research} or \url{www.youtube.com/channel/UC5-C9NqNTK39D71xK4A_0vA}.}. As shown in Section \ref{sec:dyn}, we remember that the only known dynamics in the control law is the main gravity term, i.e.: $\vec{f}_{RTN} = \begin{bmatrix}  -\frac{\mu}{r^2} & 0 & 0\end{bmatrix}^\text{T} $. 

\begin{figure}[!htb]
\centering
\subfloat[Uncontrolled orbit]{\includegraphics[width=.33\textwidth]{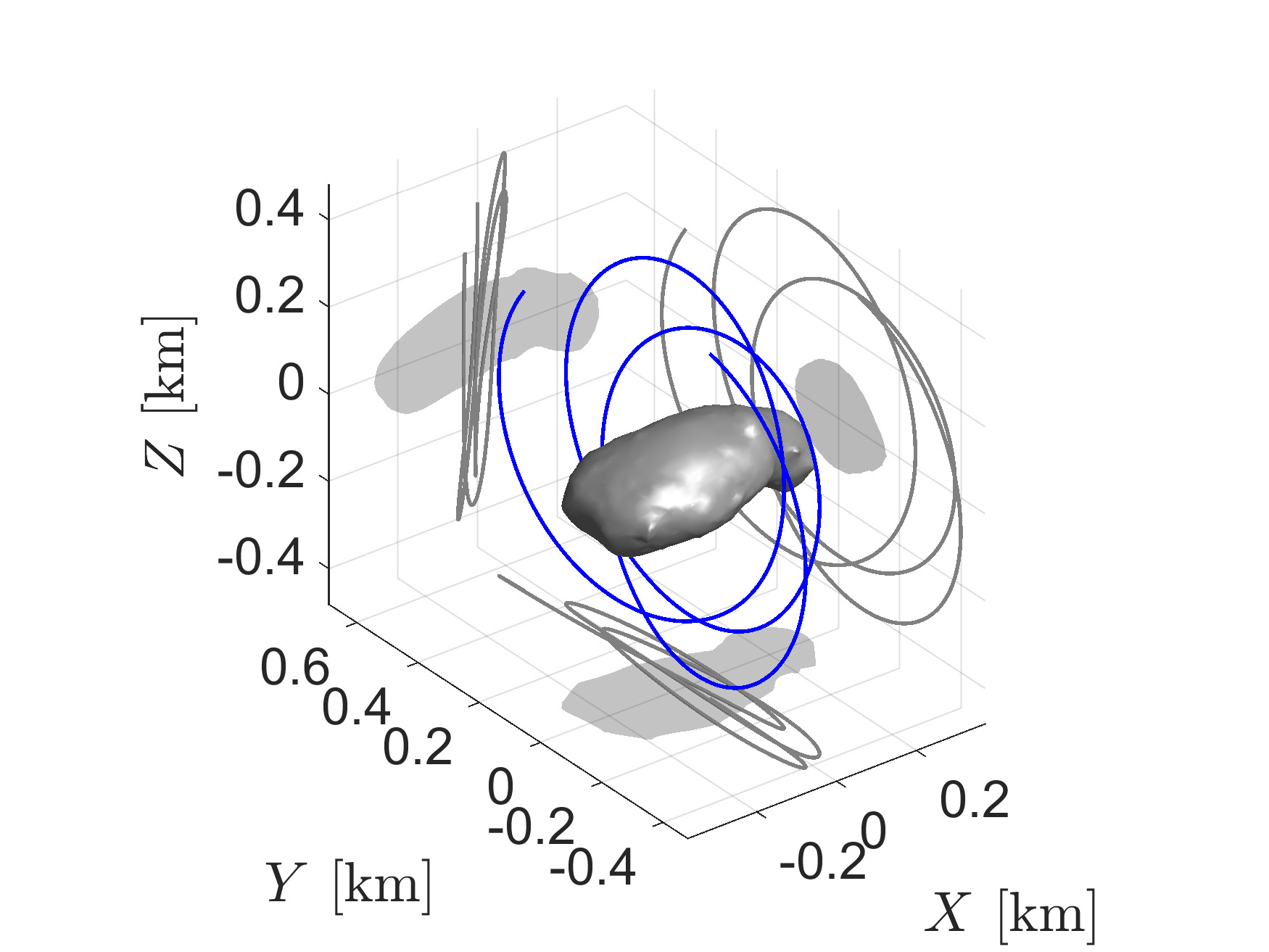}\label{Fig3a}}
\subfloat[Controlled orbit]{\includegraphics[width=.33\textwidth]{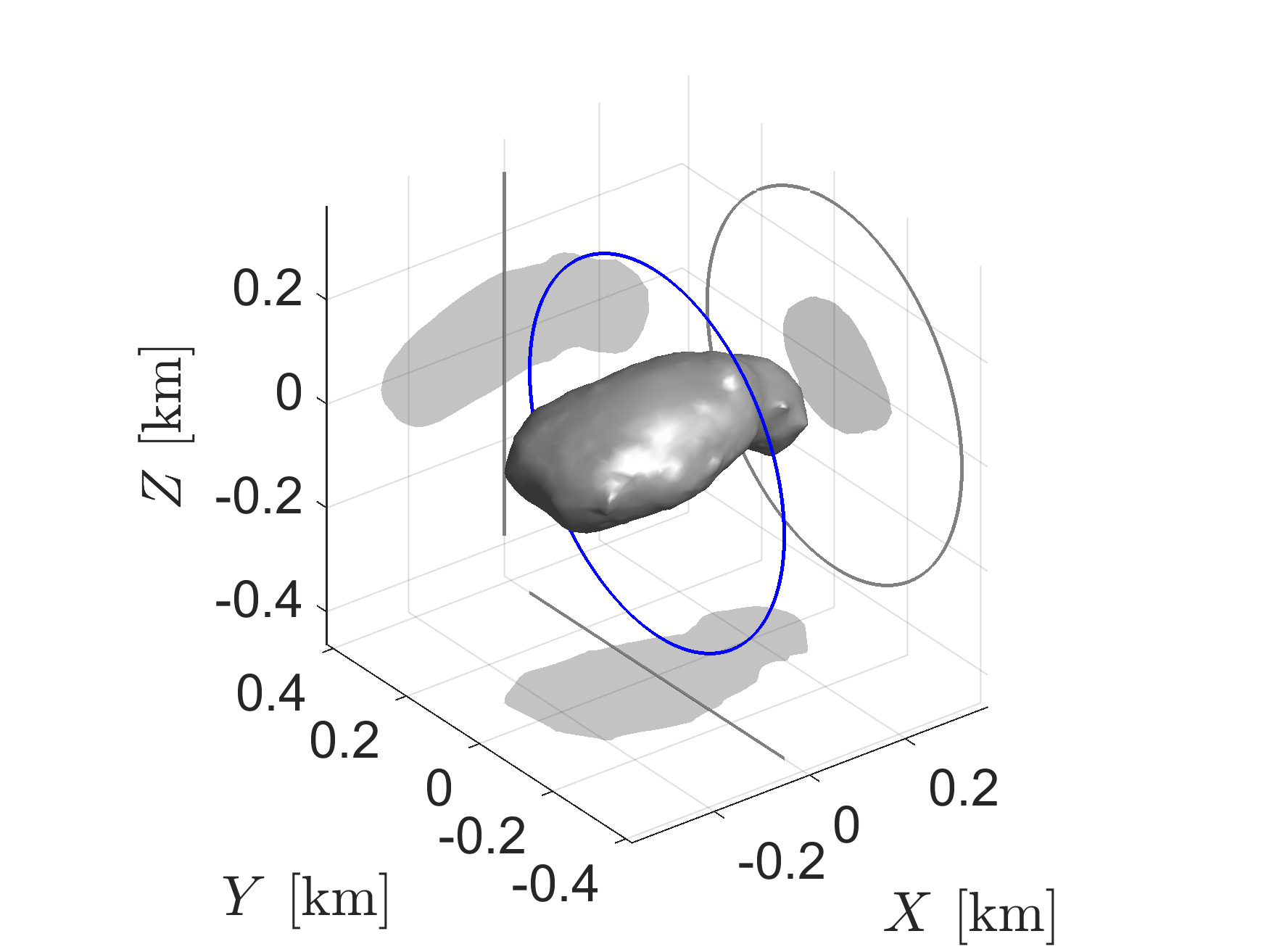}\label{Fig3b}} 
\subfloat[Control commands]{\includegraphics[width=.33\textwidth]{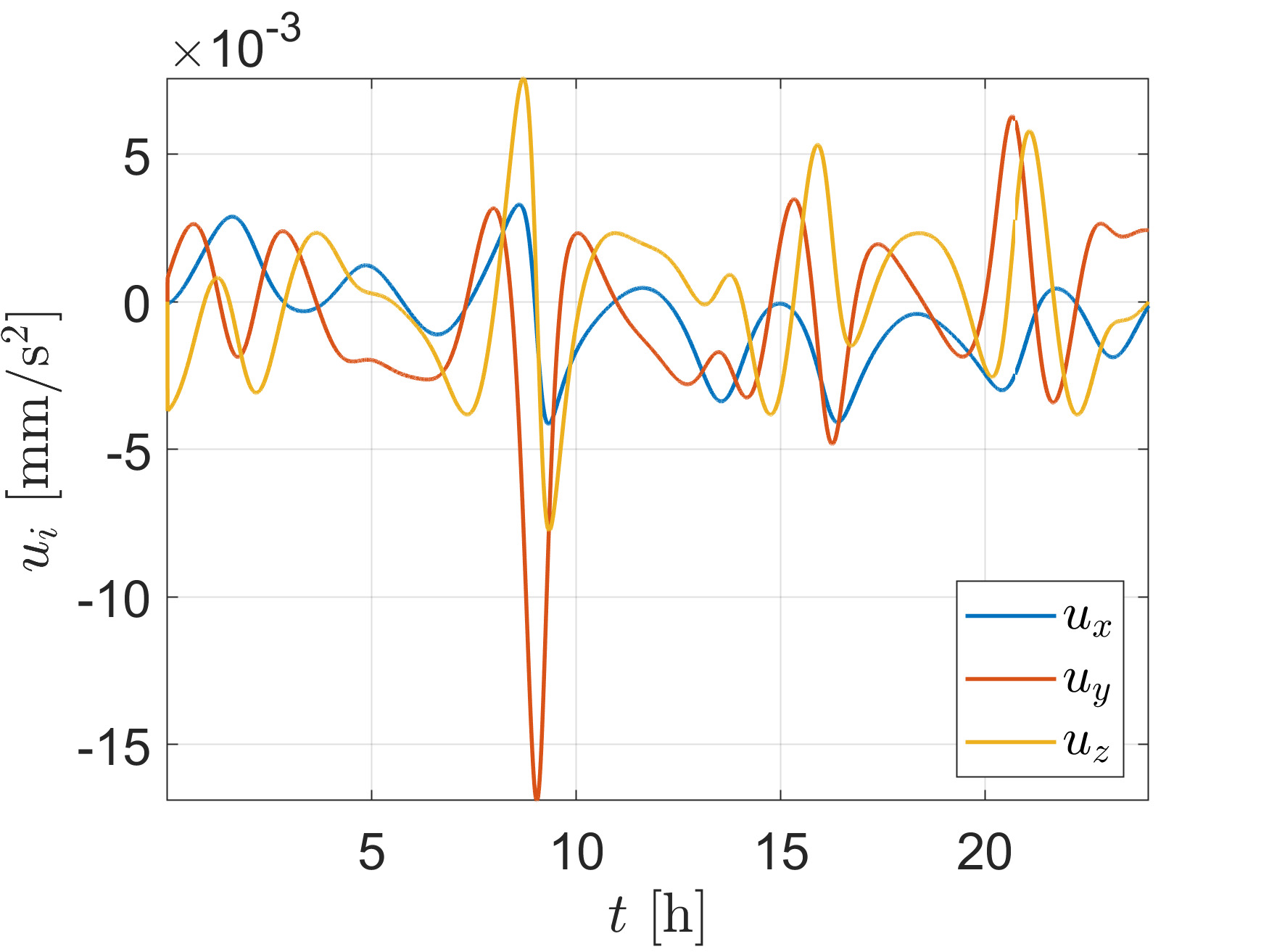}\label{Fig3c}} \\
\subfloat[Semi-major axis]{\includegraphics[width=.33\textwidth]{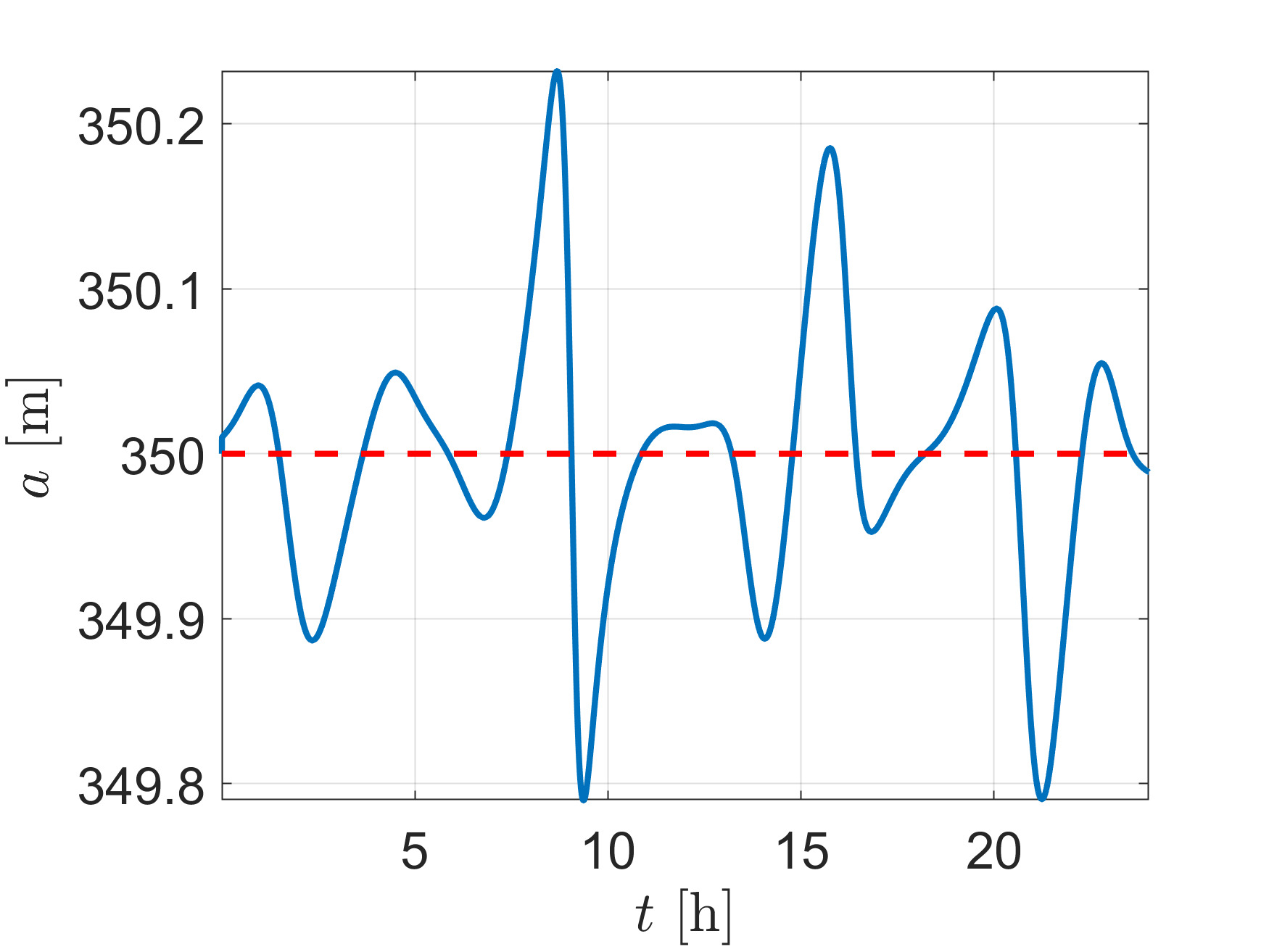}\label{Fig3d}}
\subfloat[Eccentricity]{\includegraphics[width=.33\textwidth]{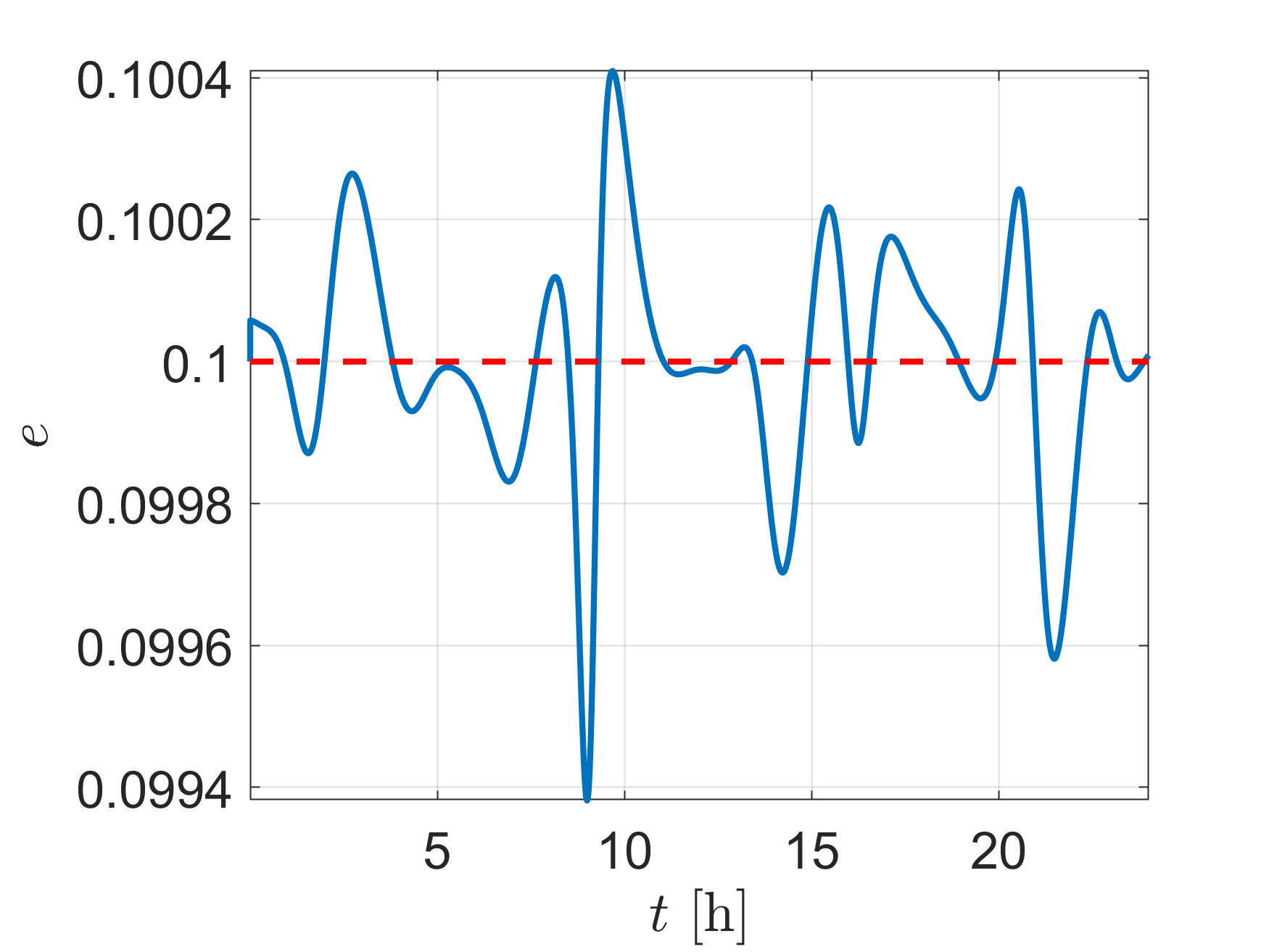}\label{Fig3e}}
\subfloat[Angular elements errors]{\includegraphics[width=.33\textwidth]{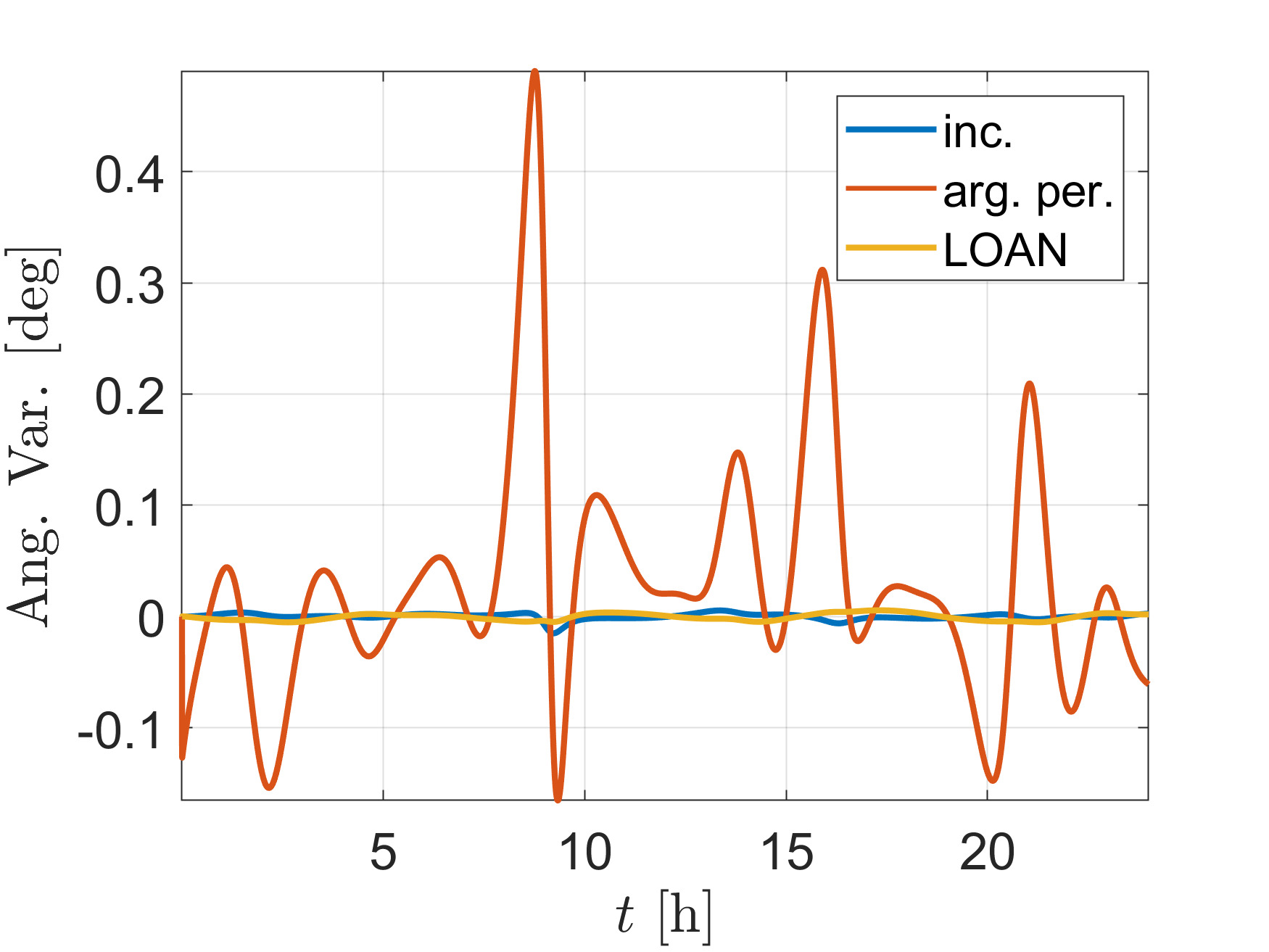}\label{Fig3f}}

\caption{Itokawa example in the inertial frame.}
\label{Fig3}
\end{figure}

We choose for our first example the asteroid Itokawa ($M=3.51 \times 10^{10}$ kg, $\nu=1.4386\times10^{-4}$ rad/s) and apply Eq. [\ref{eq:u_sat}] to control a sun-terminator orbit (i.e., angular momentum vector parallel to asteroid-sun line) with the following geometrical orbital elements: $a_d=350$ m, $e_d=0.1$, $i_d=90\degree$, $\omega_d=90\degree$, and $\Omega_d=90\degree$. The SRP parameters are (Eq. [\ref{eq:SRP}]): $S=1.695$ AU, $B_{sc}=20$ kg/m$^2$ and $\rho=1$. The control parameters are presented in Table \ref{tab:Control} as ``Itokawa'' example. Figure \ref{Fig3} shows the results for 24 hours simulation in the inertial frame. As shown in Fig. \ref{Fig3b}, the orbit is successfully maintained. If no control input is applied, the spacecraft would have a highly unstable orbit as shown in Fig. \ref{Fig3a}. One can note in Fig. \ref{Fig3c}, which represents the control commands in the Cartesian coordinates, that the saturation function is successful in avoiding chattering. Although the saturation function degrades the control performance, the error for the orbit-keeping application is very low, as shown in Figs. \ref{Fig3d} to \ref{Fig3f}, which represents each geometrical orbital element. For instance, the error for the semi-major axis is within $30$ cm, and the angular elements are all below $0.5\degree$, as seen in Fig. \ref{Fig3f}. The total $\Delta V$ for this 24 hours operation is only $0.3236$ m/s, which is reasonably low considering this close proximity orbit, with close approaches with the lobes of the rotating Itokawa.

\begin{table}[htbp]
	\fontsize{10}{10}\selectfont
    \caption{Parameters of the control for each example.}
   \label{tab:Control}
        \centering 
   \begin{tabular}{c | c | c | c | c | c} % Column formatting, 
           Example & 
      $D$ [m/s$^2$]        & 
     $n_\Phi$      & 
     $\lambda$      & 
      $\vec{s}^+$  & 
      $\vec{s}^-$    \\
      \hline
      Itokawa & $1\times 10^{-4}$ & 5 & 2 & - & -   \\
      67P & $1\times 10^{-2}$ & 5 & 2 & - & -   \\
      Bennu-2h & $1\times 10^{-2}$ & 5 & 2 & $\begin{bmatrix} 0.1 & 2.0 & 0.15  \end{bmatrix}^\text{T}$ & $\frac{1}{3}\vec{\Phi}$   \\
      Bennu-tight & $1\times 10^{-2}$ & 5 & 2 & $\begin{bmatrix} 0.02 & 0.7 & 0.05  \end{bmatrix}^\text{T}$ & $\frac{1}{3}\vec{\Phi}$   \\
      Bennu-loose & $1\times 10^{-2}$ & 5 & 2 & $\begin{bmatrix} 0.1 & 2.0 & 0.15  \end{bmatrix}^\text{T}$ & $\frac{1}{3}\vec{\Phi}$   \\
      Bennu-PWPF & $1\times 10^{-2}$ & 5 & 2 & - & -   \\
      Bennu-hyperbolic & $1\times 10^{-2}$ & 5 & 2 & $\begin{bmatrix} 0.1 & 2.0 & 0.15  \end{bmatrix}^\text{T}$ & $\frac{1}{3}\vec{\Phi}$   \\
      Bennu-Hohmann & $1\times 10^{-2}$ & 5 & 2 & $\begin{bmatrix} 0.1 & 2.0 & 0.15  \end{bmatrix}^\text{T}$ & $\frac{1}{3}\vec{\Phi}$   \\
      Bennu-Monte Carlo & $1\times 10^{-2}$ & 5 & 2 & $\begin{bmatrix} 0.1 & 2.0 & 0.15  \end{bmatrix}^\text{T}$ & $\frac{1}{3}\vec{\Phi}$   \\
   \end{tabular}
\end{table}

\begin{figure}[!htb]
\centering
\subfloat[Controlled orbit]{\includegraphics[width=.33\textwidth]{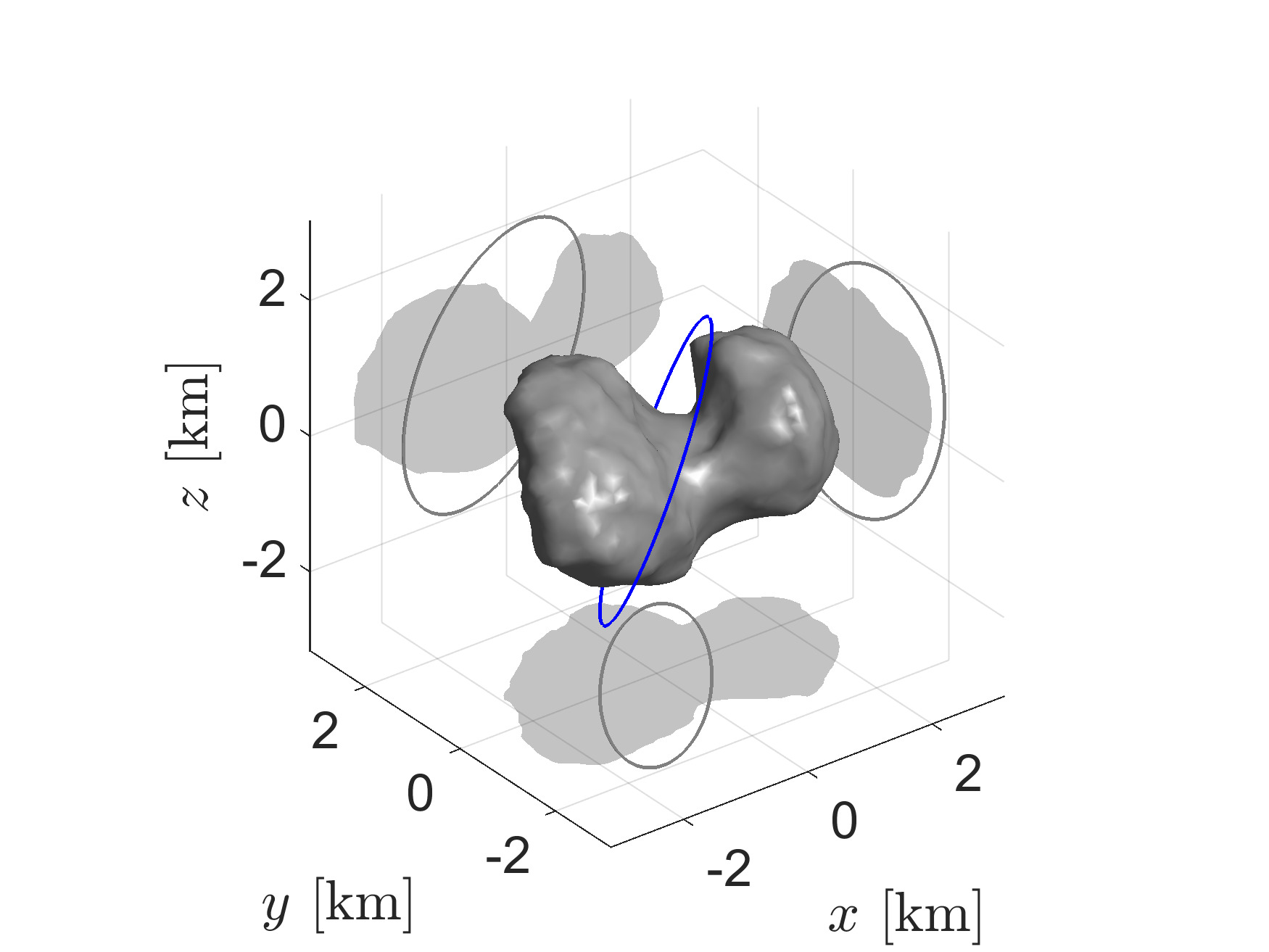}\label{Fig4a}} 
\subfloat[Control commands]{\includegraphics[width=.33\textwidth]{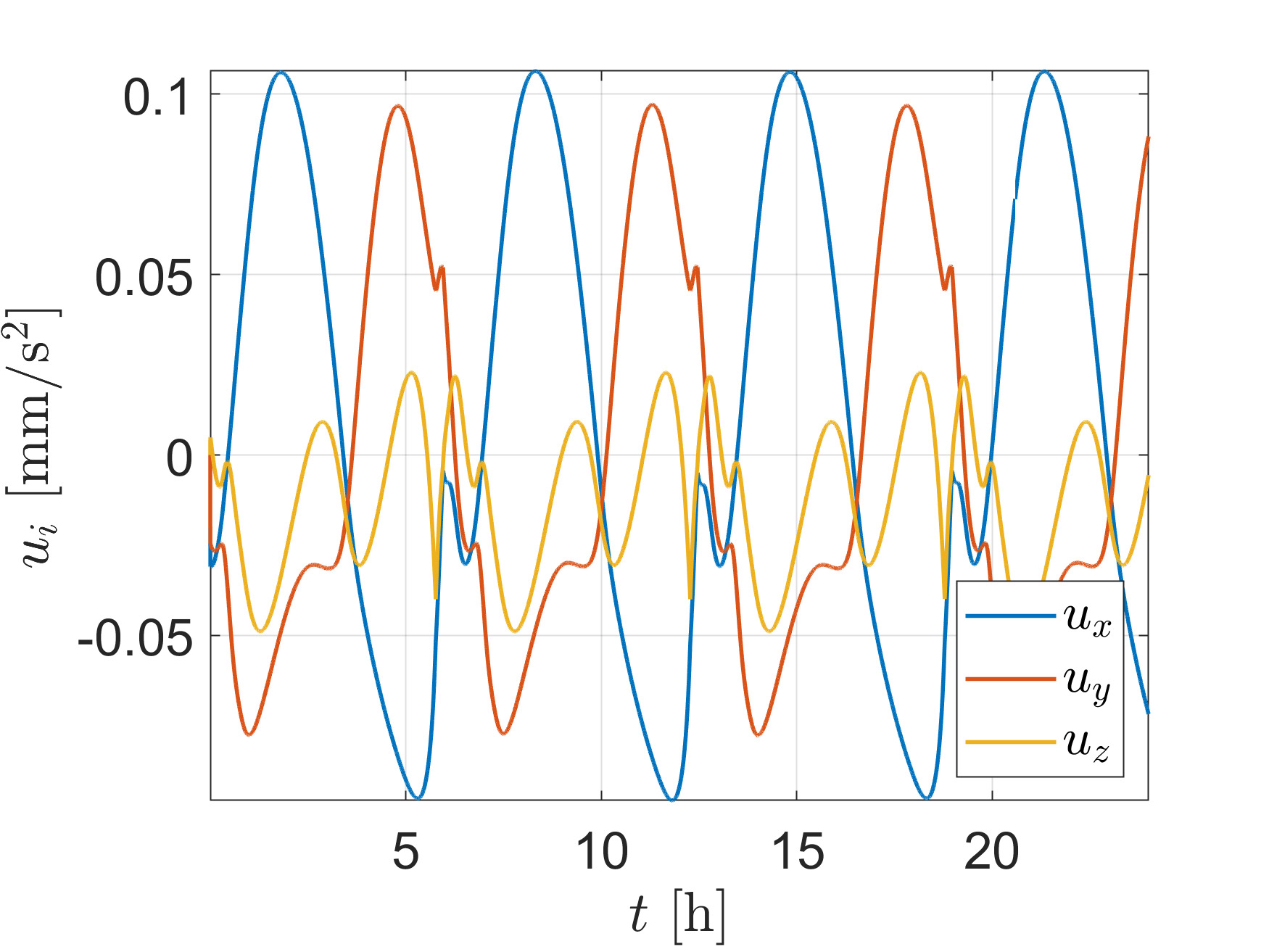}\label{Fig4b}} 

\caption{67PChuryumov–Gerasimenko example in the body-fixed frame.}
\label{Fig4}
\end{figure}

Our next example uses the comet 67PChuryumov–Gerasimenko ($M=9.982 \times 10^{12}$ kg, $\nu=1.4070\times10^{-4}$ rad/s), Fig. \ref{Fig4}, and cast the problem now as a moving path-following control \cite{negri2020path}. We choose to control the spacecraft in the body-fixed frame, which means to impose an artificial orbit, with the following geometrical orbital elements: $a_d=2100$ m, $e_d=0.15$, $i_d=110\degree$, $\omega_d=0\degree$, and $\Omega_d=50\degree$. By artificial orbit, we mean that such orbit in the body-fixed frame is impossible because the system is rotating. So a two-body problem solution does not apply, but the control can successfully create it, even with the higher level of perturbations, which include the unknown rotating state, as argued in Section \ref{sec:dyn}. The SRP parameters are (Eq. [\ref{eq:SRP}]): $S=1.243$ AU, $B_{sc}=20$ kg/m$^2$ and $\rho=1$. The control parameters are presented in Table \ref{tab:Control} as ``67P'' example. Figure \ref{Fig4a} shows the artificial orbit in the body-fixed reference frame. This kind of approach can be useful when a tight close mapping of the surface of the small body is required, which would be infeasible with no active control. Figure \ref{Fig4b}, representing the control commands, shows that the saturation function is successful in avoiding control chattering. The total $\Delta V$ for maintaining this artificial orbit is $7.0903$ m/s. For the sake of comparison, considering only the main gravity term, the necessary $\Delta V$ for maintaining a hovering on the comet's north pole (i.e., no need to cancel out accelerations due to the comet's rotation) in a distance equivalent to the artificial orbit apoapsis (2,415 m), is of 9.8633 m/s for a 24 hours maintenance. Note that this hovering $\Delta V$ would be fairly higher if considering all the disturbances, the asteroid's spin state, and transitions between hovering stations to make a hypothetical map similar to the one that the artificial orbit could make. That indicates not only the advantage that the control law can present in terms of scientific outcome but also in terms of fuel savings.

%At this point, the reader might notice that the control commands in Figs. \ref{Fig3c} and \ref{Fig4b} are in the order of micrometers per second squared. This level of acceleration command can be well executed with current technology. For instance, the Hayabusa-2 mission 28 mN ( TSUDA - System design of the Hayabusa 2-asteroid sample return mission to 1999 JU3 )

\subsection{Parametric analysis}
\label{sec:parametric}

The impact of each parameter in the sliding mode control is well documented in the literature, and it does not differ much from our control law. Here we present a brief parametric analysis so the reader can understand the magnitude of the parameters involved and how they impact our specific application. In the case of the need for a detailed near-optimal choice of the parameters and the trade-offs in an actual scenario application (control frequency, structural resonant modes, measurement rate, and others), we refer the reader to Slotine \& Li \cite{slotine1991applied} and Utkin et al. \cite{utkin2017sliding}.

\begin{figure}[!htb]
\centering
\subfloat[$n_\Phi = 5$ and $\lambda=2$]{\includegraphics[width=.33\textwidth]{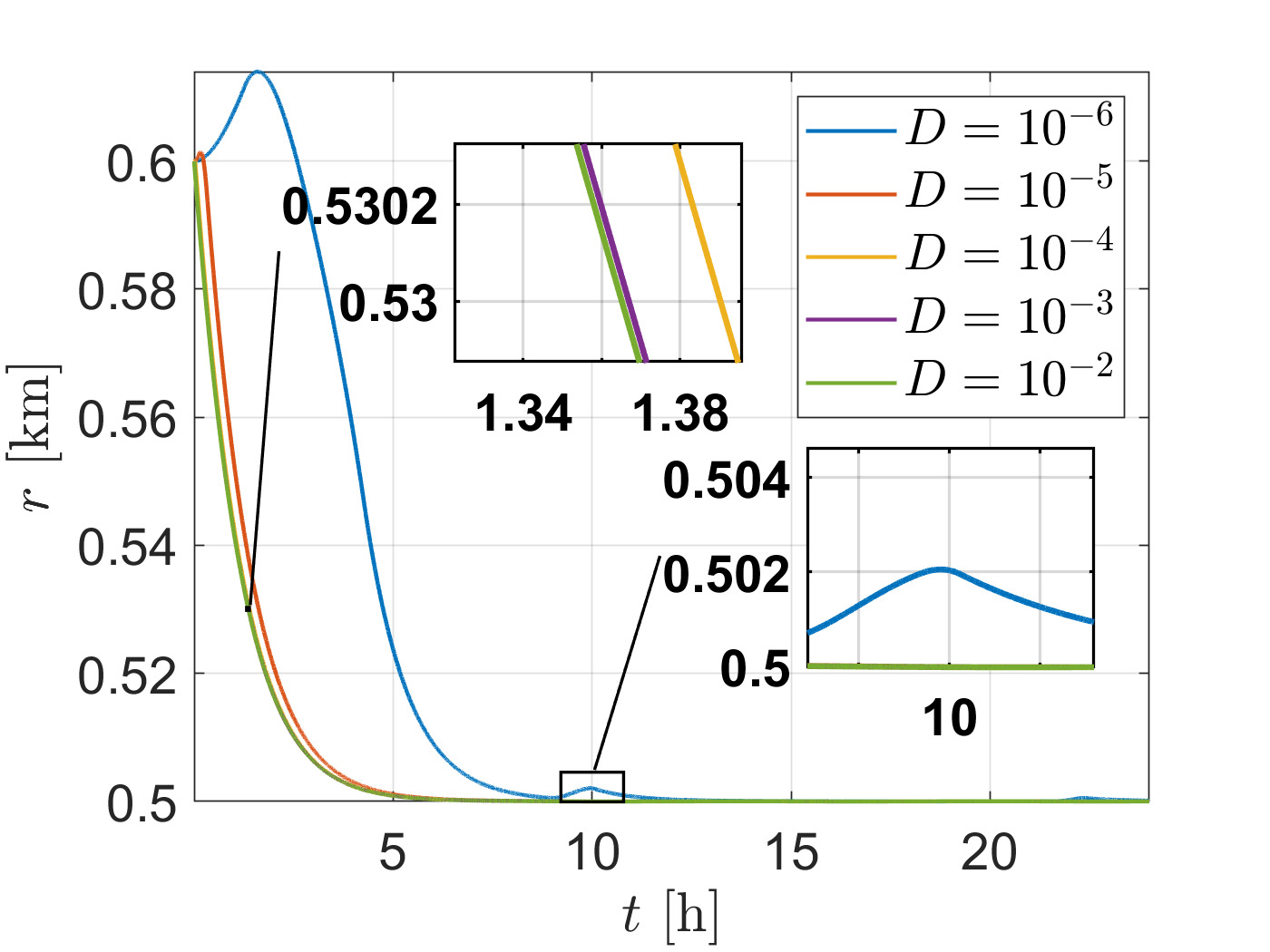}\label{Fig5a}}
\subfloat[$D=1 \times 10^{-4}$ and $n_\Phi = 5$]{\includegraphics[width=.33\textwidth]{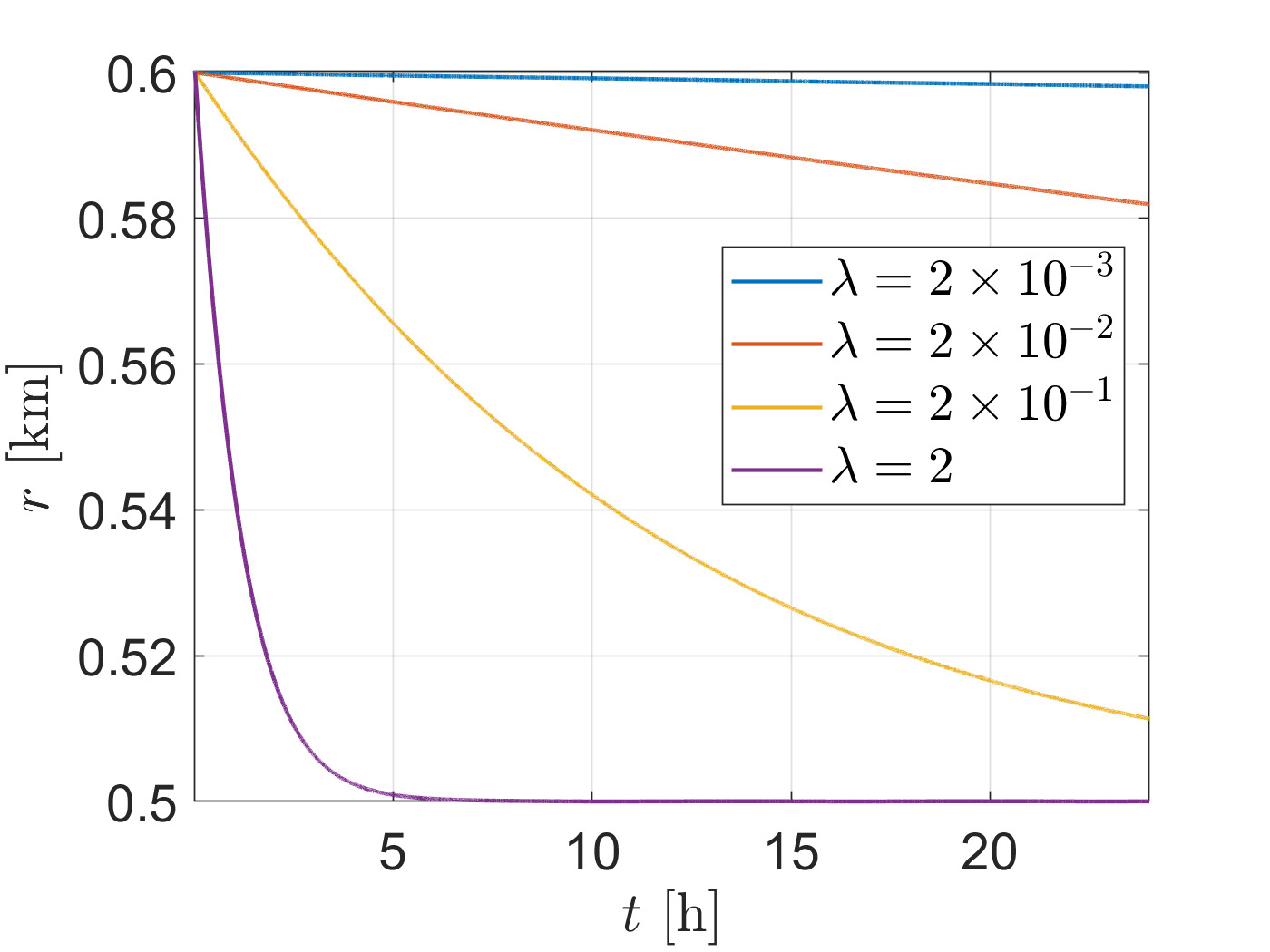}\label{Fig5b}}
\subfloat[$D=1 \times 10^{-4}$ and $\lambda=2$]{\includegraphics[width=.33\textwidth]{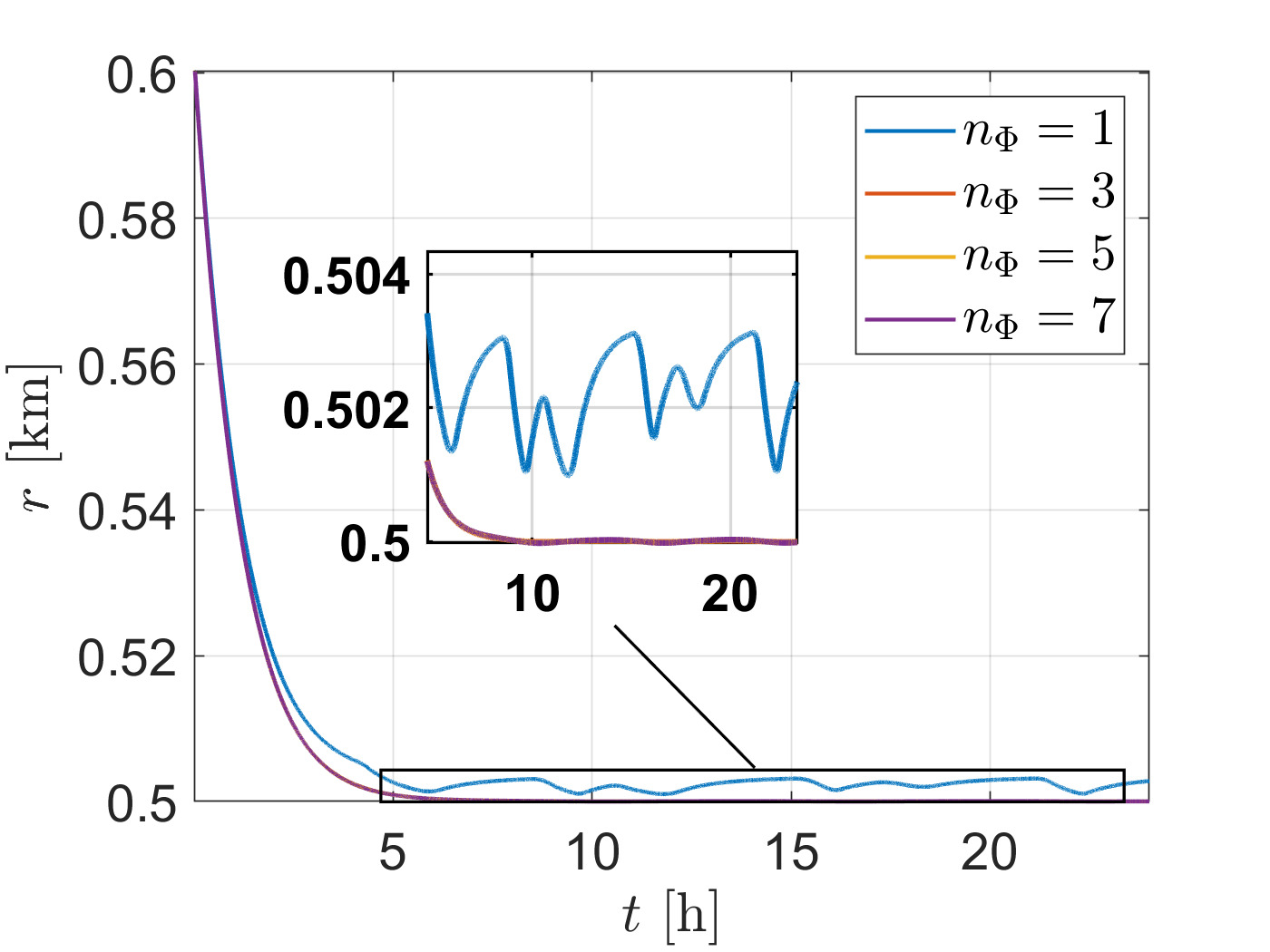}\label{Fig5c}}  %\\
%\subfloat[Total $\Delta V$, for $n_\Phi = 5$ and $\lambda=2$]{\includegraphics[width=.33\textwidth]{Fig5d}\label{Fig5d}}
%\subfloat[Total $\Delta V$, for $D=1 \times 10^{-4}$ and $\lambda=2$]{\includegraphics[width=.33\textwidth]{Fig5e}\label{Fig5e}}
%\subfloat[Total $\Delta V$, for $D=1 \times 10^{-4}$ and $n_\Phi = 5$]{\includegraphics[width=.33\textwidth]{Fig5f}\label{Fig5f}}

\caption{Distance from asteroid's barycenter for different control parameters, circular sun-terminator orbit about Itokawa, in the inertial frame.}
\label{Fig5}
\end{figure}

Figure \ref{Fig5} shows the results when varying the three main parameters of the control, the assumed disturbance level $D=D_R=D_T=D_N$ (directly associated with the gain matrix $K$), the size of boundary about the sliding surface, summarized here in the parameter $n_\Phi$ ($\vec{\Phi}= n_\Phi \text{diag}(K)$), and the sliding surface constant $\lambda$. In this analysis, we use the spherical harmonics model for simulating the asteroid's gravity field. A circular sun-terminator orbit around Itokawa with $500$ m radius is chosen to be controlled. We assume that the initial condition of the spacecraft is 600 m from the asteroid's center of mass at its north pole. In order to summarize the parametric analysis of the orbital control in a single figure, we choose to represent the performance using the distance of the spacecraft to the asteroid's barycenter $r$. 

The $K$ matrix is directly associated with the rate of convergence to the sliding surface \cite{negri2020path,slotine1991applied}. As we are calculating it as a function of the assumed disturbance level $D$, we can note in Fig. \ref{Fig5a}, which presents the performance for different values of $D$, that those larger magnitudes of the assumed disturbance result in faster convergence as expected. The actual perturbation level for this operational condition is in the order of $1 \times 10^{-7}$ m/s$^2$. That is why for $D=1\times 10^{-6}$  m/s$^2$ some bumps are noted in the curve, reaching an error in $r$ of the order of 2 meters. These bumps correspond to moments when the spacecraft is close to the Itokawa's rotating lobes, meaning a larger perturbation level, close in magnitude to the assumed $D=1\times 10^{-6}$  m/s$^2$, which consequently degrades the control performance.

While the matrix $K$ is related to the rate of convergence to the sliding surface, the parameter $\lambda$ is intimately attached to the rate of convergence to the desired orbit once the sliding surface was already reached \cite{negri2020path,slotine1991applied}. As we can check in Fig. \ref{Fig5b}, the parameter $\lambda$ is the most impactful. For $\lambda=2$, the spacecraft converges to the orbit in 5 hours. For decreasing values of $\lambda$ the convergence is increasingly longer, with a bit more of a day for $\lambda=0.2$ and achieving tens of days for $\lambda=2 \times 10^{-3}$~\footnote{The plot is limited to 24 hours to visualize better the trend in behavior for the different $\lambda$.}.

Figure \ref{Fig5c}, representing the performance for different values of $n_\Phi$, shows a typical case of chattering. The more significant errors for $n_\Phi=1$, when compared to the other cases, is a result of the inability of the saturation function in attenuating the gain of the matrix $K$ for the chosen control frequency ($0.25$ Hz), which should be higher if a boundary layer around the sliding surface for $n_\Phi=1$ is required. The other cases of $n_\Phi$ do not present much difference in performance for this particular application, and their curves are practically equal in Fig. \ref{Fig5c}. 

\begin{figure}[!htb]
\centering
\subfloat[$\Delta t = 0.01$ s]{\includegraphics[width=.33\textwidth]{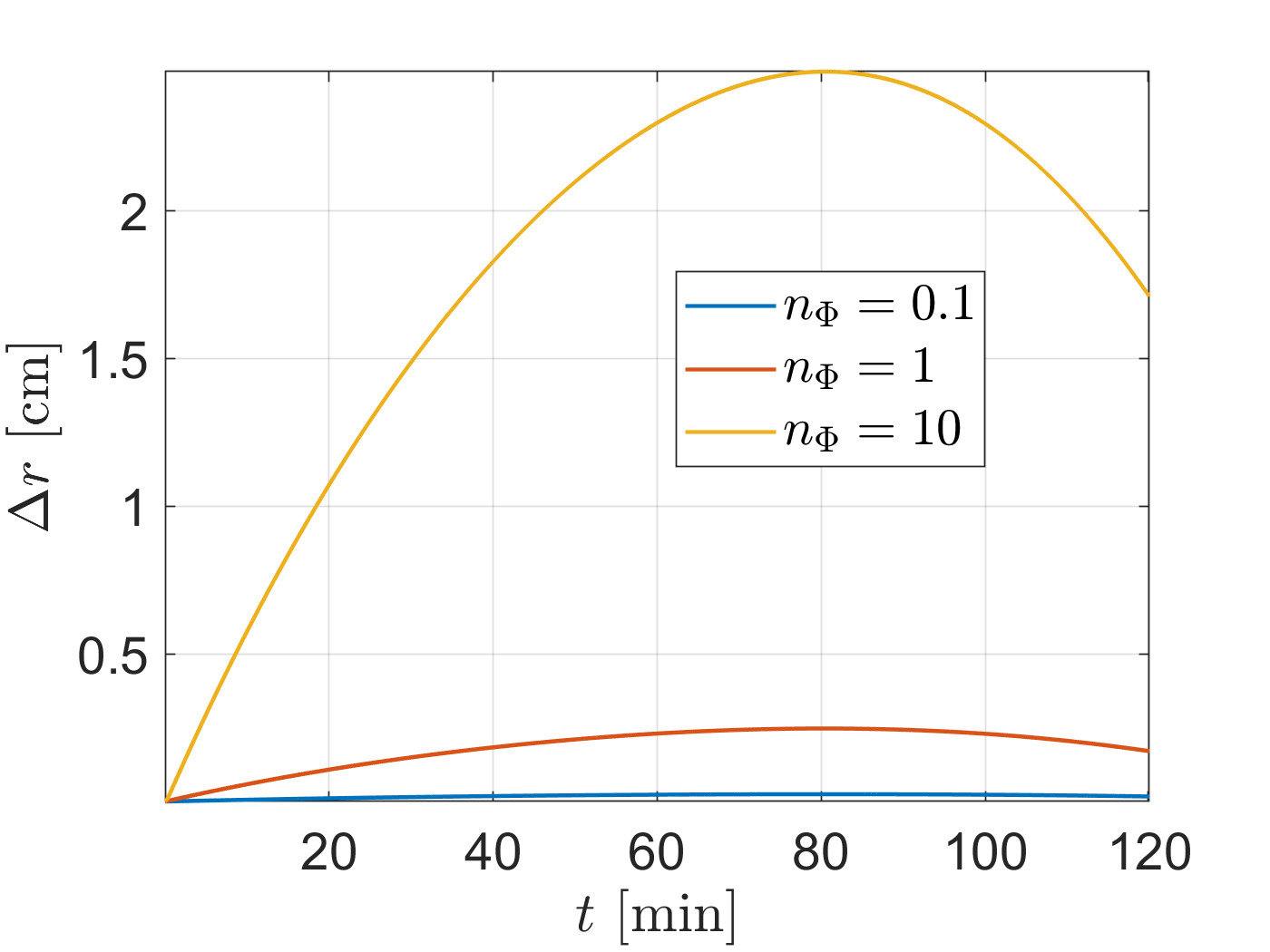}\label{Fig6a}}
\subfloat[$\Delta t = 1$ s]{\includegraphics[width=.33\textwidth]{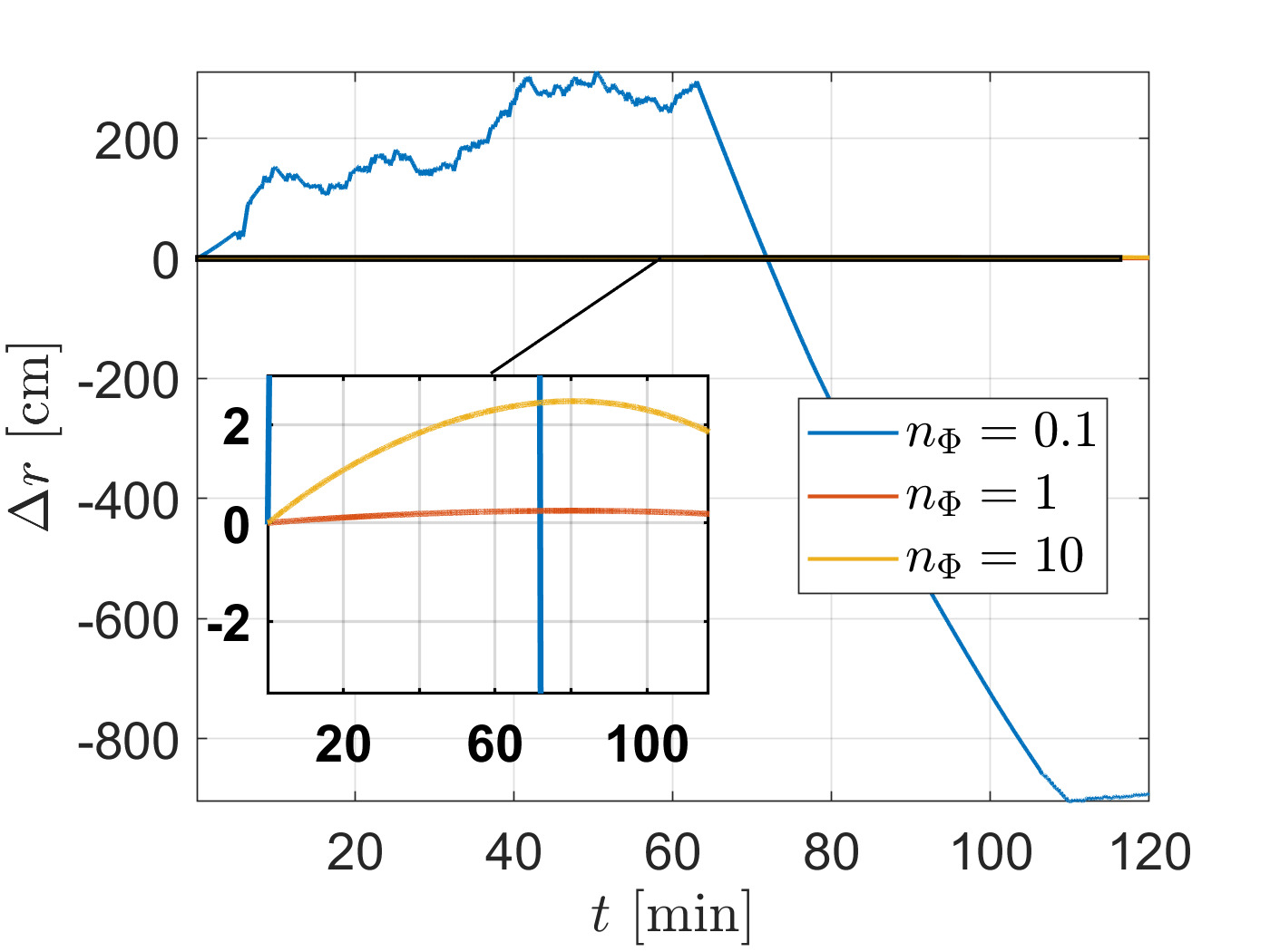}\label{Fig6b}} 
\subfloat[$\Delta t = 10$ s]{\includegraphics[width=.33\textwidth]{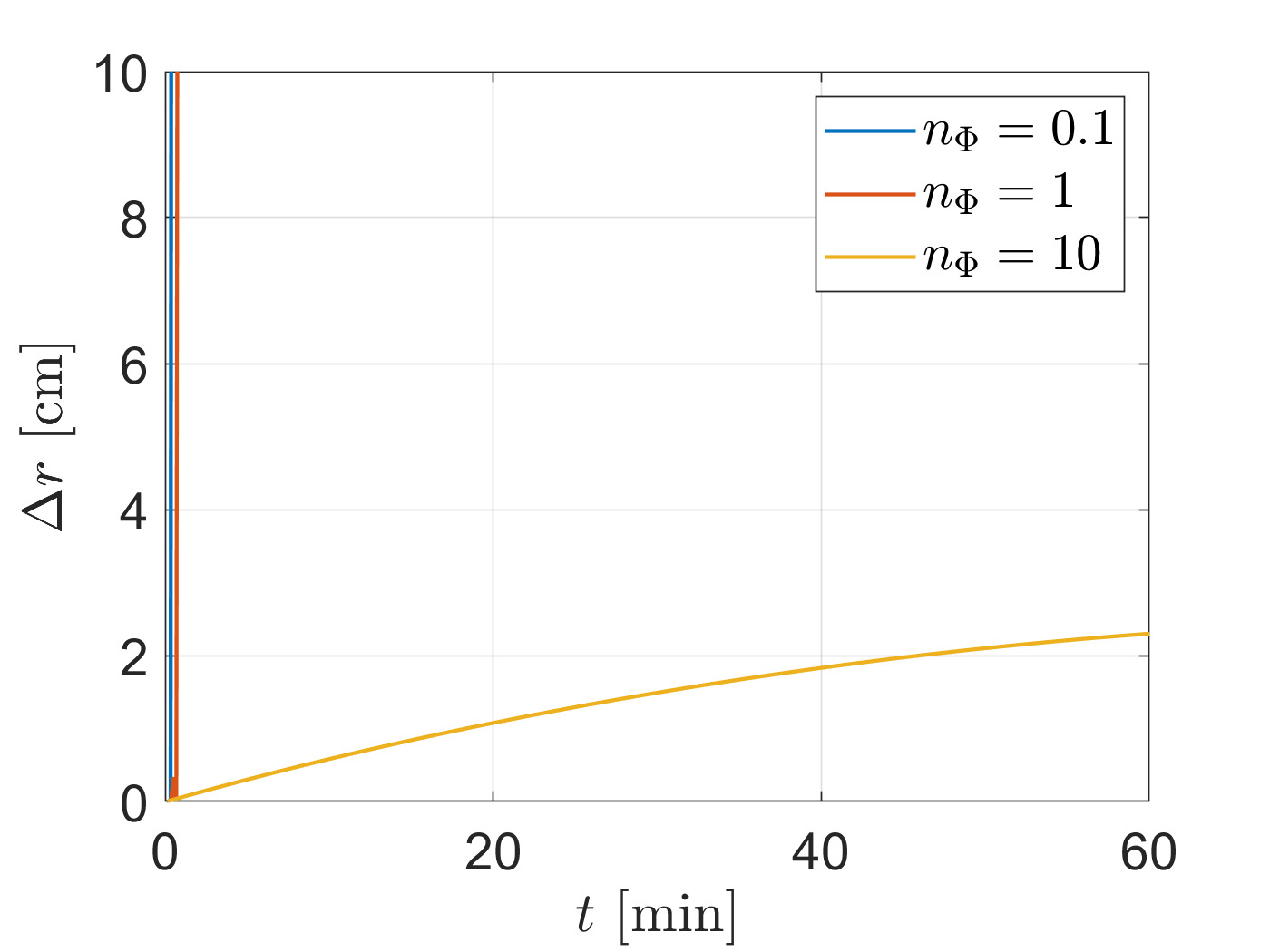}\label{Fig6c}}

\caption{Error in $r$ for different control frequencies, circular sun-terminator orbit about Itokawa, in the inertial frame, for $D=1 \times 10^{-3}$ and $\lambda=2$}
\label{Fig6}
\end{figure}

As we pointed out before, we had to limit our control frequency to 0.25 Hz to make our simulations' computational costs less demanding. Now, we show a small analysis of the impact of the control frequency in the performance of the orbital control. The necessity of the saturation function to attenuate the control gain is also intimately associated with the control frequency~\cite{utkin2017sliding}. Therefore, in Fig. \ref{Fig6}, we show different simulations considering different control frequencies for the same 500 m circular sun-terminator orbit and also consider the application of different $n_\Phi$. As one can check in Fig. \ref{Fig6a}, presenting the results considering a control update in each $\Delta t = 0.01$ s, i.e., 100 Hz, this choice of control frequency is good enough to accommodate different magnitudes of $n_\Phi$, with no indication of chattering. Increasing performance for smaller values of $n_\Phi$ can also be noted, as we should expect. Increasing the update time of the control for $\Delta t = 1$ s, Fig. \ref{Fig6b}, one can quickly note that a boundary layer of $n_\Phi = 0.1$ is unable of removing the chattering, and the system presents high-frequency oscillations (these could excite structural resonant modes) and poor performance. Finally, setting $\Delta t = 10$ s, only a boundary layer for $n_\Phi = 10$ can maintain the control performance, while the error in $r$ explodes for the others. We end this control frequency discussion by bringing the reader's attention to the fact that the proposed control law is very promising for operating in real-time. It is entirely analytical~\footnote{We remember that the matrix $F$ to be inverted is a $3\times3$ matrix, and its analytical inversion is easy to obtain~\cite{negri2020path}.}, handling all of the system disturbances without the need of solving a complex dynamical model.

\subsection{Bennu scenario analysis}

Considering how the orbit-keeping law could operate, we will have a more practical approach to the problem in the following examples. We now also consider noise in the measurements, and thus we will make all subsequent analysis focused on the asteroid Bennu ($M=7.329 \times 10^{10}$ kg, $\nu=4.0684\times10^{-4}$ rad/s), with SRP parameters $S=0.8969$ AU, $B_{sc}=20$ kg/m$^2$ and $\rho=1$, currently visited by the OSIRIS-Rex mission, in order to apply a realistic noise level. For this sake, the noise level at the data cutoff (DCO) of Orbit-B Phase presented in Williams et al. \cite{williams2018osiris} is considered. They separate the noise level in each RTN component. However, we will consider the most significant magnitude of the noise at the DCO rounded up, so we consider an additive white Gaussian noise with $1\sigma$ dispersion in each Cartesian component of the position and velocity vectors, respectively, as $\sigma_{\vec{r}}=0.8$ m and $\sigma_{\vec{v}}=1\times10^{-4}$ m/s. We also apply in most of the following examples the control in Eq. [\ref{eq:u_sat_hys}] (we will note when it is not the case), noting that a realistic orbit-keeping control will not work uninterruptedly in most of the cases (i.e., with the propulsive system constantly applying a thrust). It is also assumed that the propulsive system is limited in delivering the exact control command calculated by the control law. That is done by considering a $1\sigma$ dispersion of 3$\%$ in each component of the calculated control command. We note that the thrusters' uncertainty assumption is very conservative, as in reality, considering different thrusters, they tend to remain below a 1$\sigma$ of 1\% \cite{tajmar2004indium,machuca2019autonomous,antreasian2019early,snyder2019electric},

As a final restriction added to the system, the maximum acceleration command is assumed to be limited to 1 mm/s$^2$. We note that this limitation to the control command is only to set an additional degree of challenge to the control law. It is not an intrinsic limitation attached to the control law or an autonomous mission. That variable is more related to mission goals and the choice of hardware, which is beyond the scope of this engineering note. Of course, if a highly precise orbital station-keeping is needed, low thrust engines that deliver the same acceleration level of the system's perturbations should be employed. Note that this can go as far as using colloidal micro-Newton thrusters, like the ones used in the LISA Pathfinder~\cite{racca2010lisa}. Conversely, if a coarser orbit-keeping is enough, hydrazine engines in the order of a few Newtons are enough. It is a matter of mission trade-offs.

The OSIRIS-REx Orbit-B Phase is an initially circular sun-terminator orbit with a semi-major axis of 1 km, which will become a slightly eccentric and back to circular on the order of months \cite{williams2018osiris}. Scheeres et al. \cite{scheeres2013design} show that if the orbit's semi-major axis is reduced to 500 m, the orbit dynamics becomes more complex, and the spacecraft collides with the asteroid in less than 100 days. So, in our first simulation, we control this 500 m circular orbit considering the hysteresis of the control switcher. The control parameters are presented in Table \ref{tab:Control} as ``Bennu-2h'' example. In this first application, we consider a measurement update every 2 hours, which is compatible with the OSIRIS-REx mission measurement frequency~\cite{williams2018osiris,takahashi2021autonomous}, and corrupted by the noise we defined before. It is considered as if the spacecraft has an onboard integrator to propagate, between the measurements, the navigation solution up to the next update. In order to stress the feasibility of our proposition, we consider that the propagator only considers the central body term of the gravity field and the calculated control command~\footnote{Note that, as we defined before, the actual applied control has a 3$\%$ 1$\sigma$ dispersion around the calculated control.}. The trajectory that is solved in real-time onboard, with the initial state from the navigation solution, is assumed as the nominal trajectory.

\begin{figure}[!htb]
\centering
\subfloat[Controlled orbit]{\includegraphics[width=.33\textwidth]{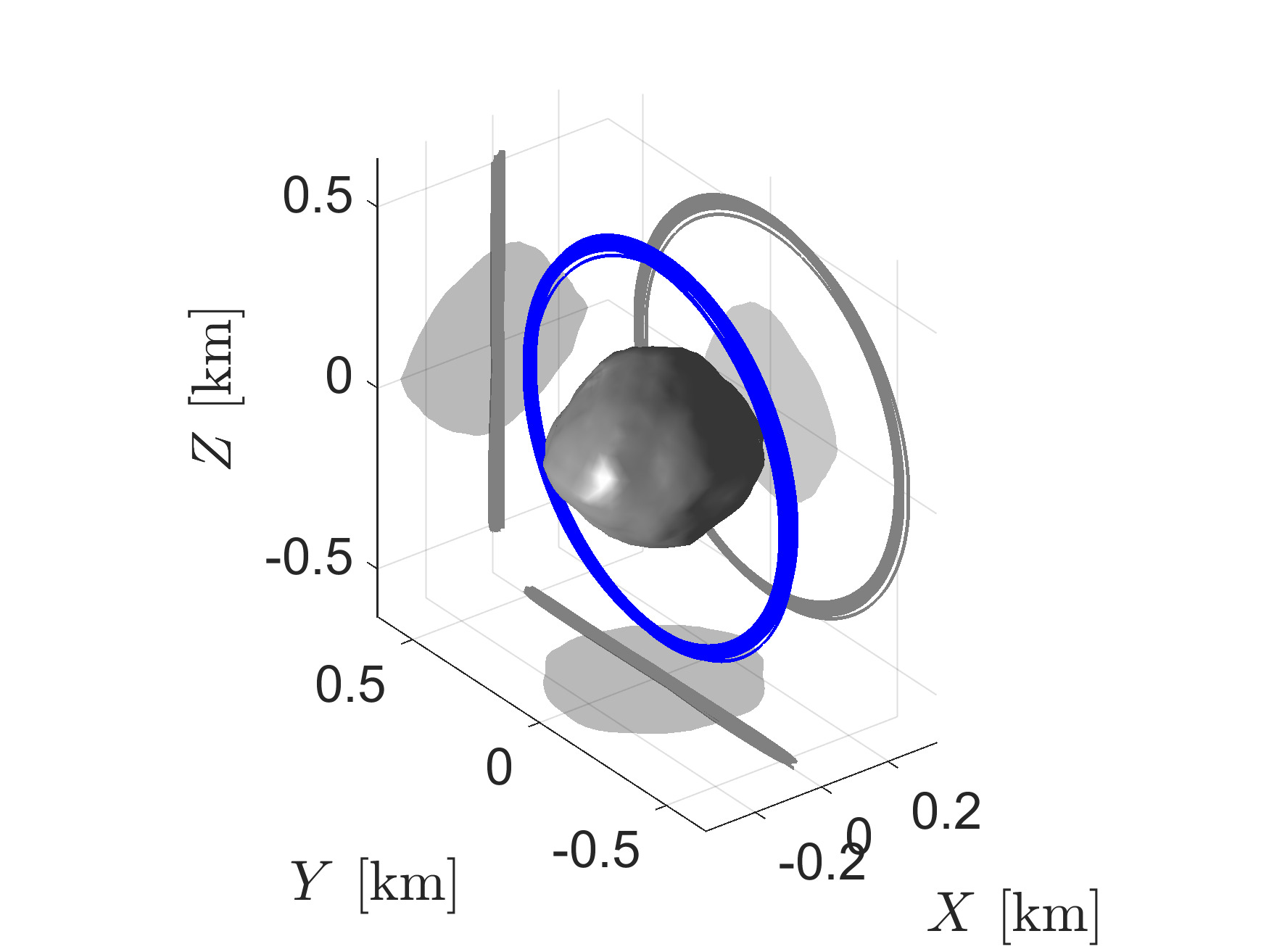}\label{Fig7a}} 
\subfloat[Control commands]{\includegraphics[width=.33\textwidth]{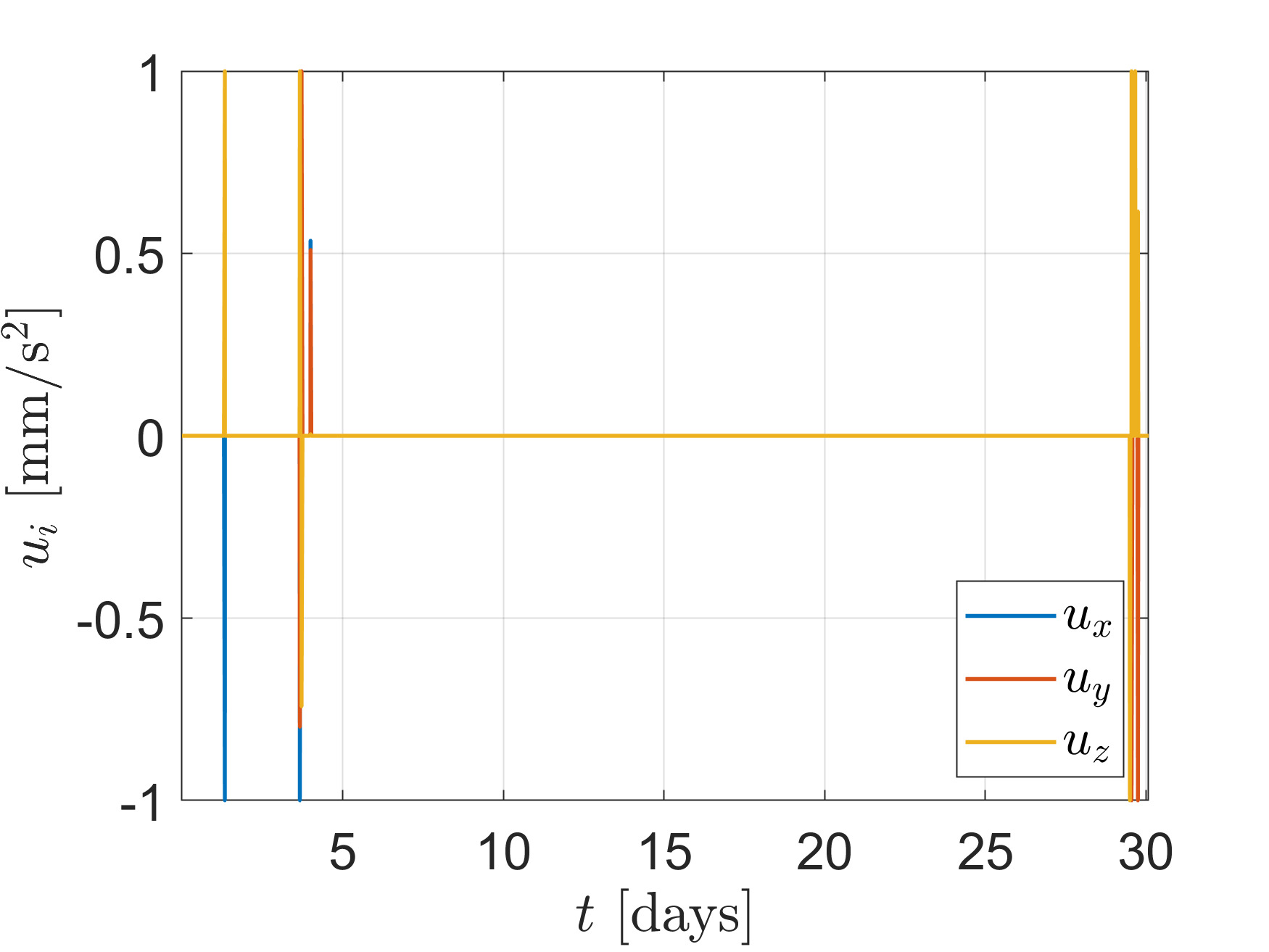}\label{Fig7b}}  

\caption{Bennu example in the inertial frame, for 2 hour navigation solution update.}
\label{Fig7}
\end{figure}

Figure \ref{Fig7} shows the controlled orbit for that case, for an operation time of one month. Because the simulation time is long, we apply the spherical harmonics model for representing the asteroid's gravity field. As we can check in Fig. \ref{Fig7a}, the trajectory is successfully controlled with our proposition. The control commands in Fig. \ref{Fig7b} show that the control switcher is successful in its task, with only a few burns in the whole operation. The total $\Delta V$ in this 30 days operation is only 4.95 cm/s, which is around 0.23$\%$ of the 22 m/s budget $\Delta V$ estimated for the two-year Bennu proximity operations of the OSIRIS-REx mission \cite{williams2018osiris}. We ask the reader's attention to the fact that no meticulous orbital analysis is applied to design the control switcher, nor in the choice of the control parameters. Furthermore, many of the assumed limitations we impose on the scenario (only the main gravity term is known by the control law and the onboard integrator, high dispersion in the executed control commands, noise level, and others) can be considered exaggerated. That serves to emphasize the advantage that our orbital control can present for small body missions.

We showed that the proposed control is feasible if an autonomous operation is applied with a measurement rate similar to the OSIRIS-REx. As it is of today, current missions autonomous navigation software applies a batch-sequential filtering approach for navigating the spacecraft~\cite{team2000autonomous,riedel2006autonav,bhaskaran2012autonomous}. The main reasons are the advantages of a batch approach. It allows a stochastic analysis of the data before processing and presents better stability than sequential filtering when processing sparse measurements - as it uses a consistent reference trajectory -. Moreover, in many cases, there is no demand to apply sequential filtering due to the large time constants of the application~\cite{bhaskaran2012autonomous}. That is well compatible with a measurement rate such as the one of the OSIRIS-REx~\cite{takahashi2021autonomous}. 

Nevertheless, if an approach is equally or more challenging than the ones exemplified by the Itokawa and 67PChuryumov–Gerasimenko examples, Figs. \ref{Fig3} and \ref{Fig4}, is applied, the time constants of the dynamics may decrease drastically, which in turn increases the need for a larger measurement frequency. That could result in the demand for a sequential filtering approach~\cite{ohira2020autonomous}. Therefore, because a greater frequency of measurements, and so sequential filtering, might be the only solution in specific challenging applications, we will consider the same noise we defined before adding to each control step's position and velocity. In other words, we consider that the measurement and estimation frequency is equal to the control frequency, and the uncertainty of the state's estimation is passed to the control with no smoothness. We will also choose even more severe conditions for the orbit to be controlled. We begin by considering a circular orbit with $a_d=350$ m, $i_d=45\degree$, and $\Omega_d=45\degree$. Note that this is not a sun-terminator orbit, making null the stabilizing effects of the SRP on the orbital plane (which is already small in such proximity). If this orbit evolves uncontrolled in time, it will collide with Bennu in less than 7 hours.

\begin{figure}[!htb]
\centering
\subfloat[Controlled orbit]{\includegraphics[width=.33\textwidth]{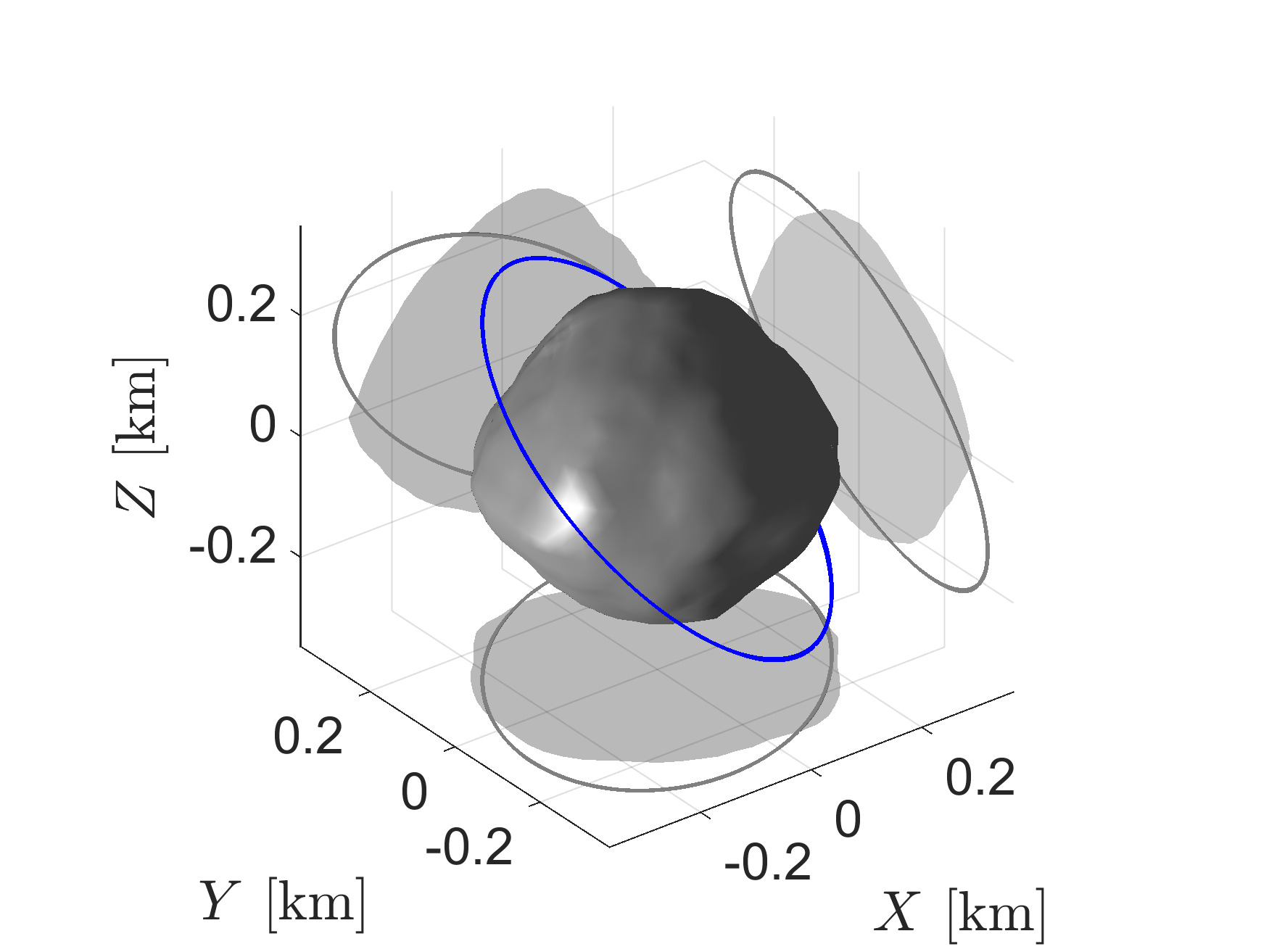}\label{Fig8a}}
\subfloat[Control commands]{\includegraphics[width=.33\textwidth]{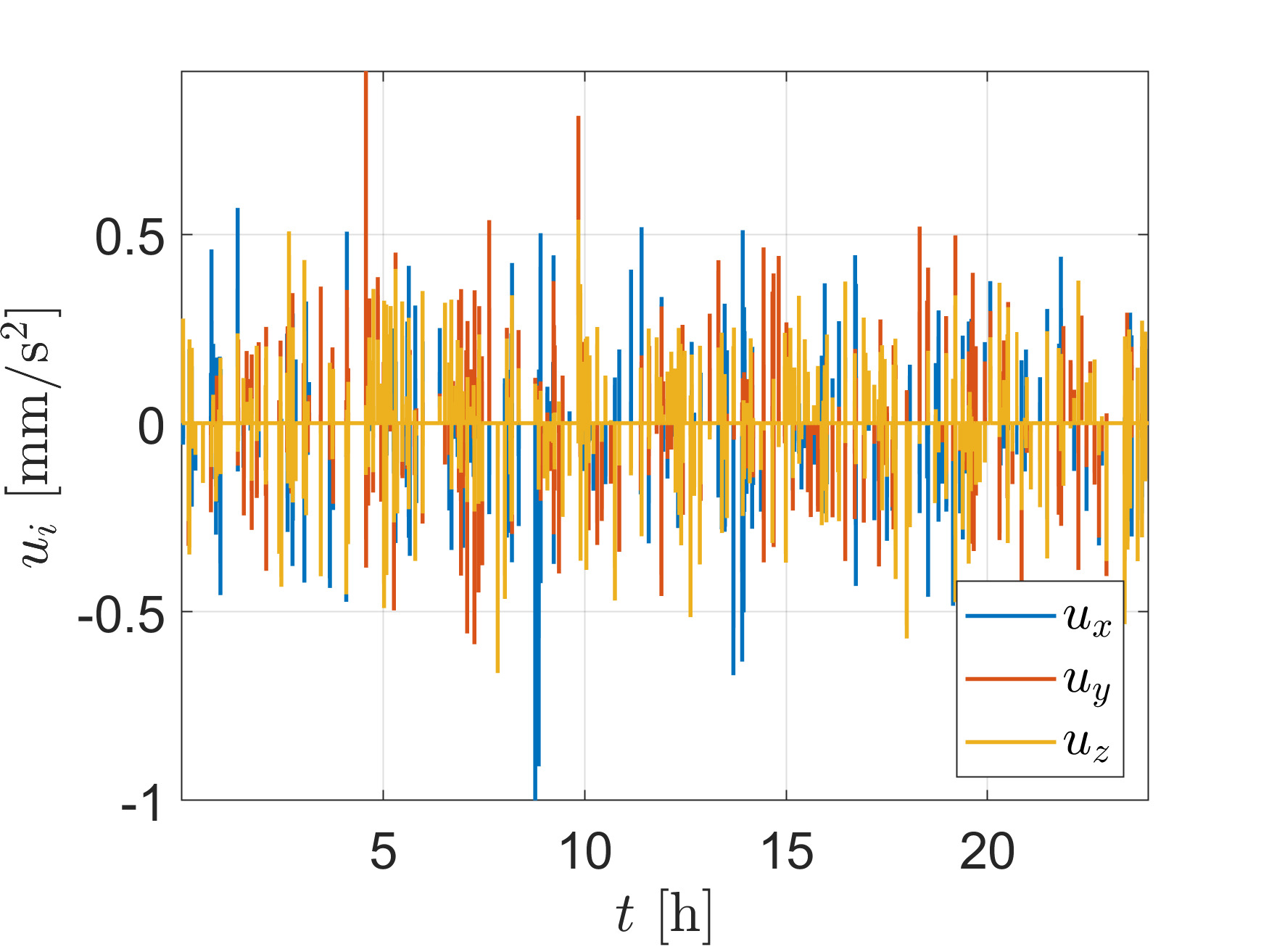}\label{Fig8b}}  \\
\subfloat[Semi-major axis]{\includegraphics[width=.33\textwidth]{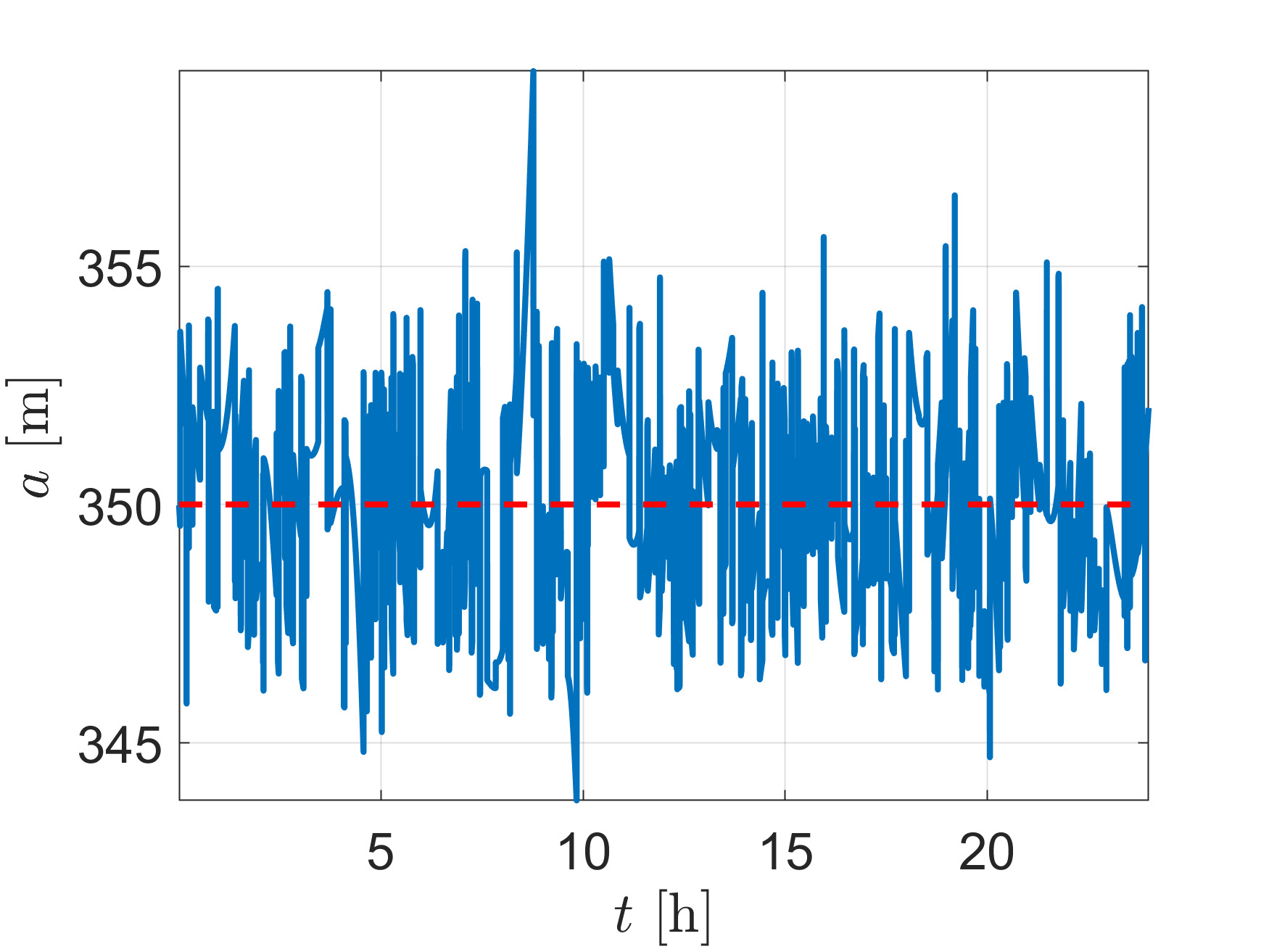}\label{Fig8c}}
\subfloat[Eccentricity]{\includegraphics[width=.33\textwidth]{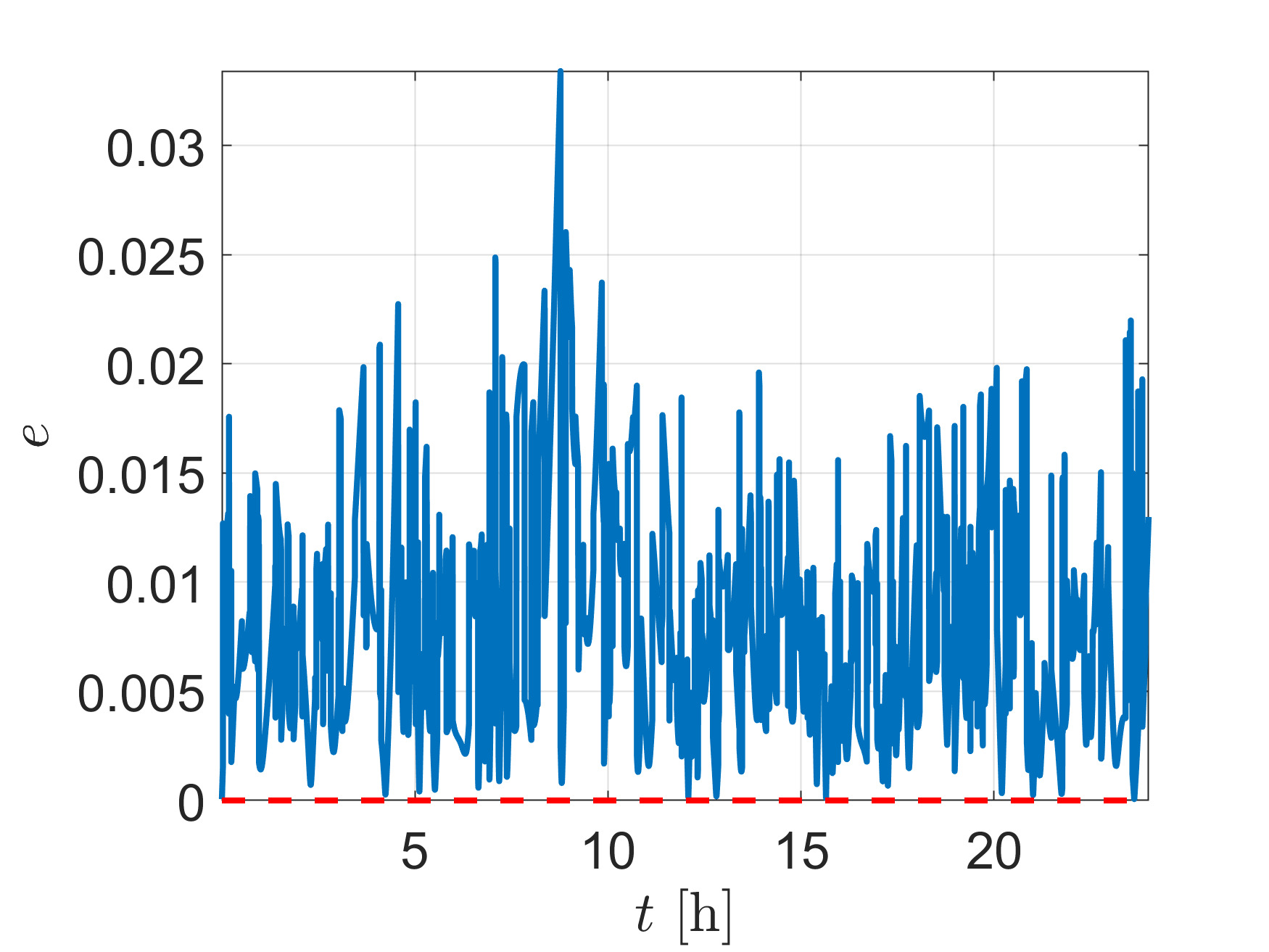}\label{Fig8d}}
\subfloat[Angular elements errors]{\includegraphics[width=.33\textwidth]{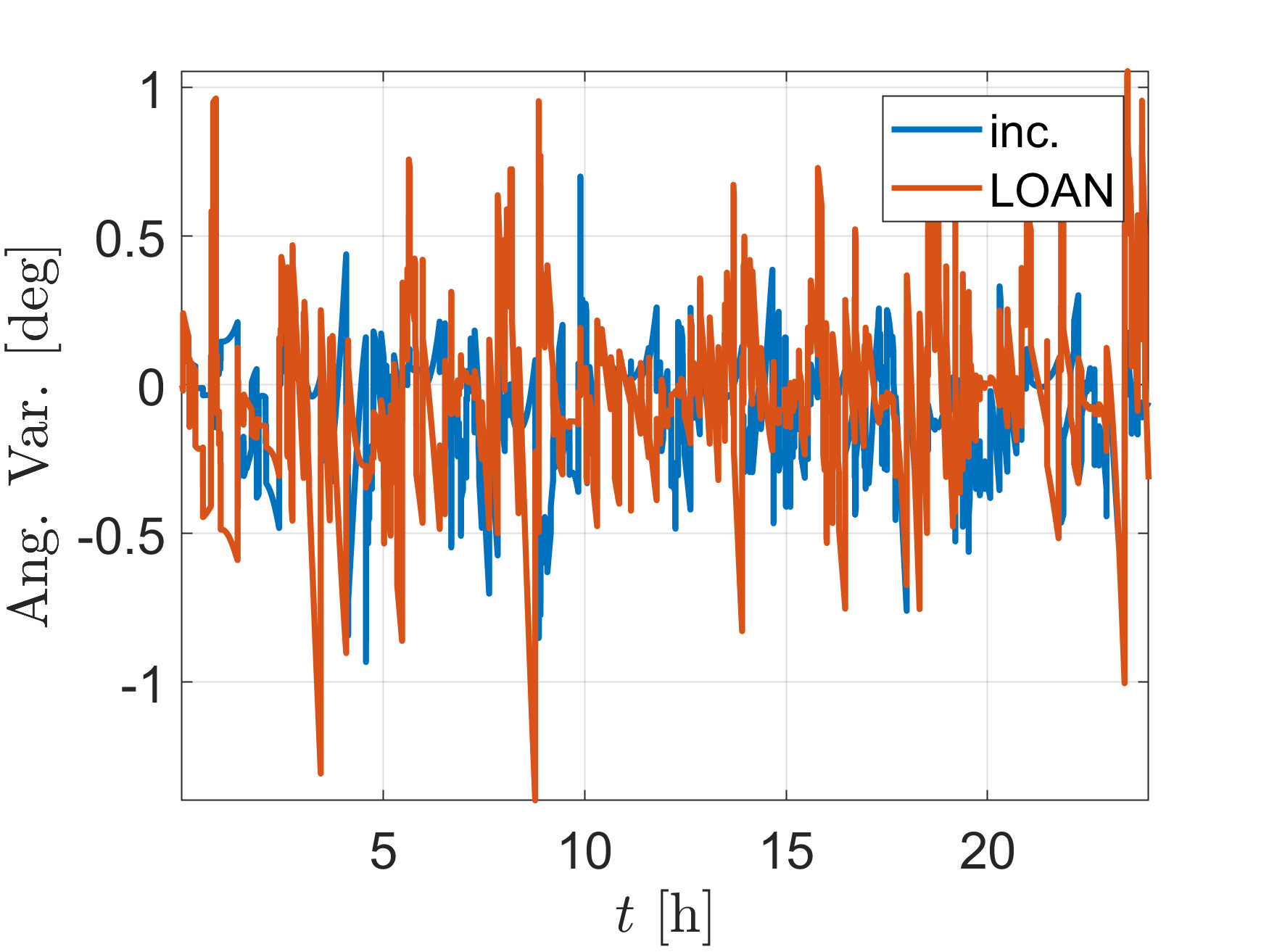}\label{Fig8e}}

\caption{Bennu tighter hysteresis example, in the inertial frame.}
\label{Fig8}
\end{figure}

Using Eq. [\ref{eq:u_sat_hys}], we consider two different cases to control this orbit. We first consider a tighter control, by adjusting the hysteresis parameters to: $\vec{s}^+=\begin{bmatrix} 0.02 & 0.7 & 0.05  \end{bmatrix}^\mathbb{T}$ and $\vec{s}^-=\frac{1}{3}\vec{\Phi}$. The other control parameters are presented in Table \ref{tab:Control} as ``Bennu-tight'' example. Figure \ref{Fig8a} shows that the orbit is successfully maintained for a 24 hours simulation, which is sufficient to complete more than four orbital periods. The control components are shown in Fig. \ref{Fig8b}. For this application, the idle-thrusters periods are in the order of minutes due to the tight hysteresis controlling the on-off switch of the control. Figures \ref{Fig8c} to \ref{Fig8e} show that the errors in the orbital elements are small even with all the practical considerations. In this one-day operation, the $\Delta V$ is 50.45 cm/s, about 2$\%$ of the $\Delta V$ estimated for the two-year OSIRIS-REx proximity operations. Therefore, if the science team decides it would be scientifically valuable to operate in such a close orbit, even though it is naturally unenviable, the proposed path-following control could safely execute it with little compromise of the mission budget $\Delta V$.

\begin{figure}[!htb]
\centering
\subfloat[Controlled orbit]{\includegraphics[width=.33\textwidth]{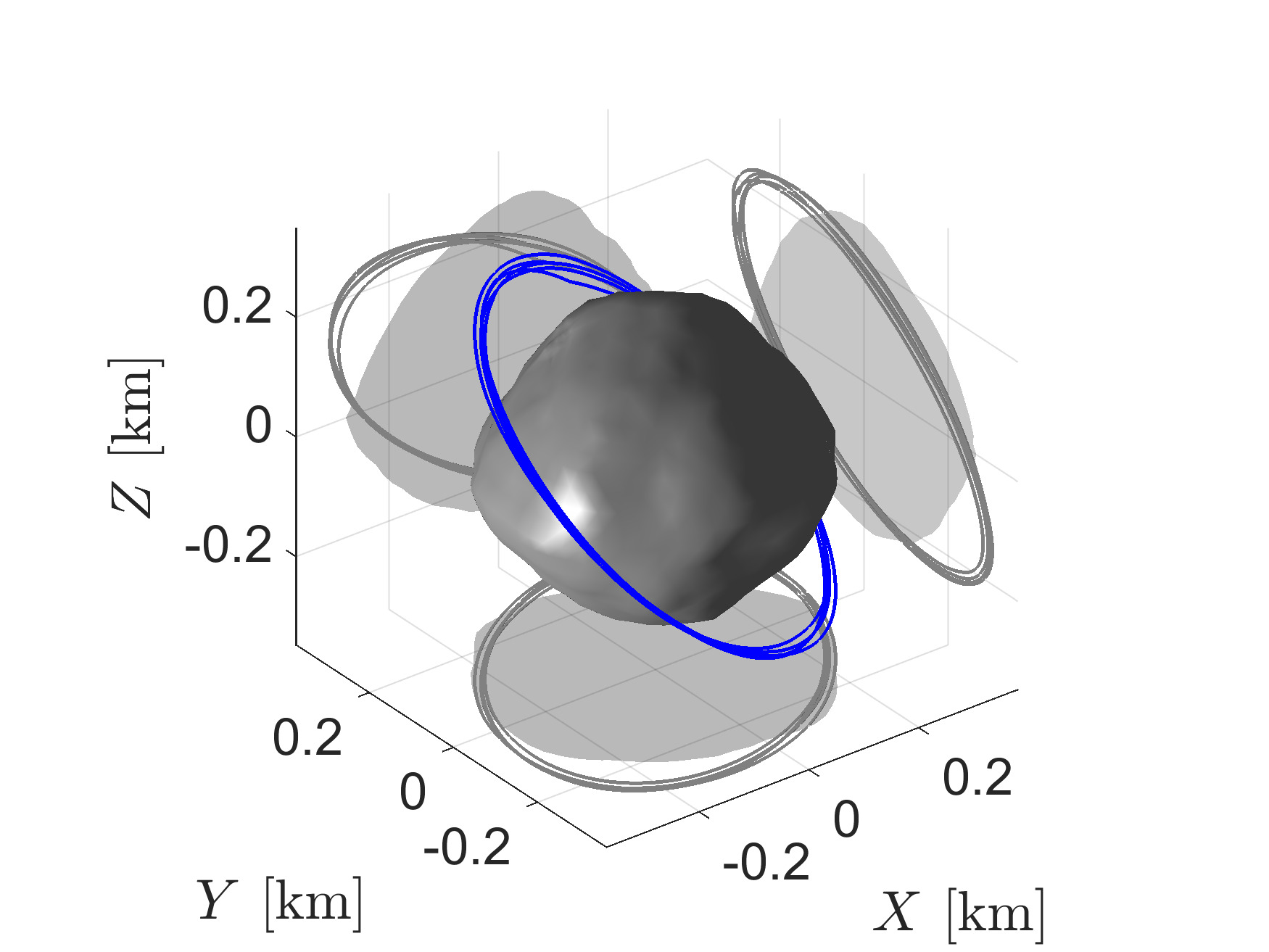}\label{Fig9a}}
\subfloat[Control commands]{\includegraphics[width=.33\textwidth]{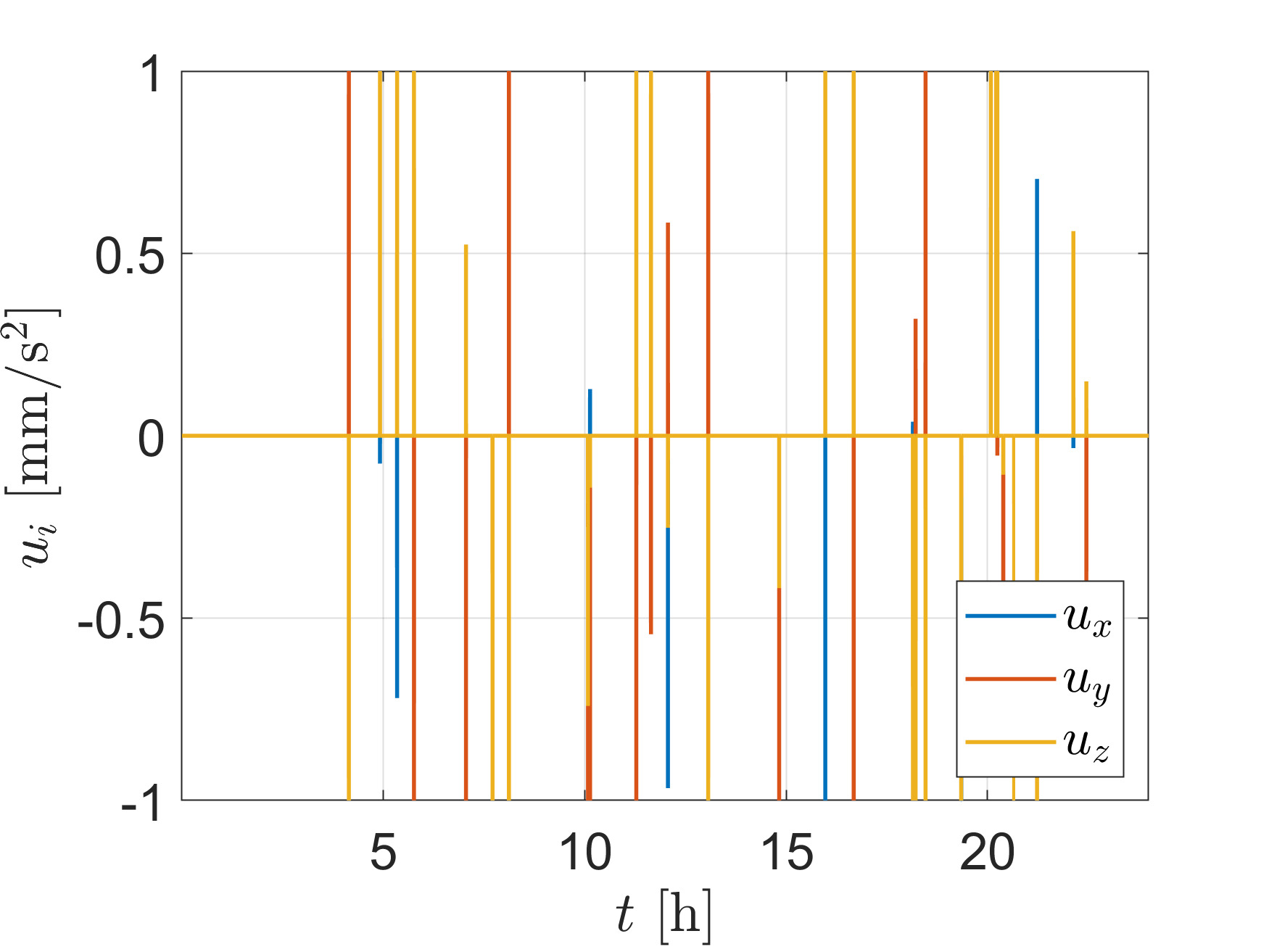}\label{Fig9b}}  \\
\subfloat[Semi-major axis]{\includegraphics[width=.33\textwidth]{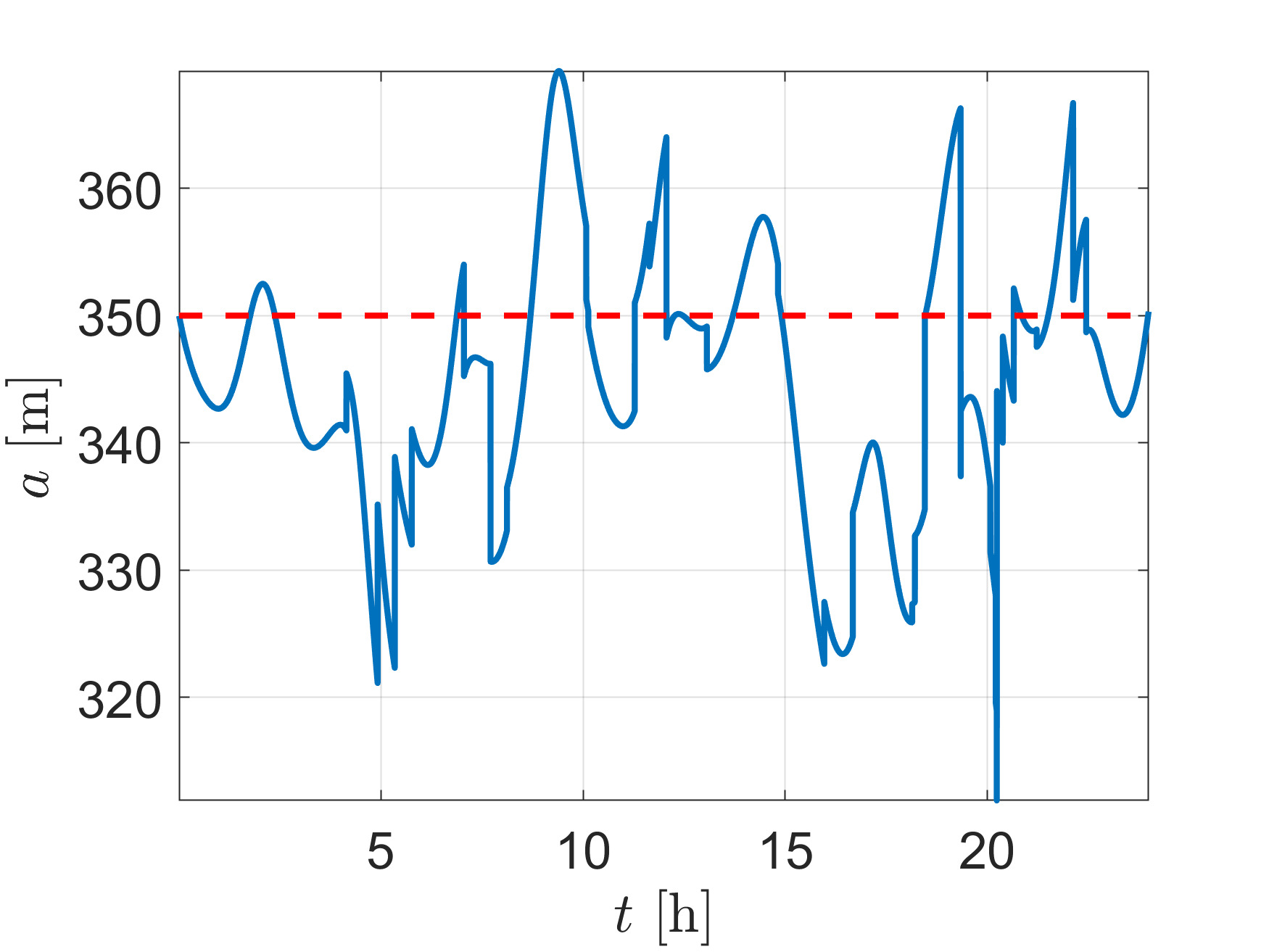}\label{Fig9c}}
\subfloat[Eccentricity]{\includegraphics[width=.33\textwidth]{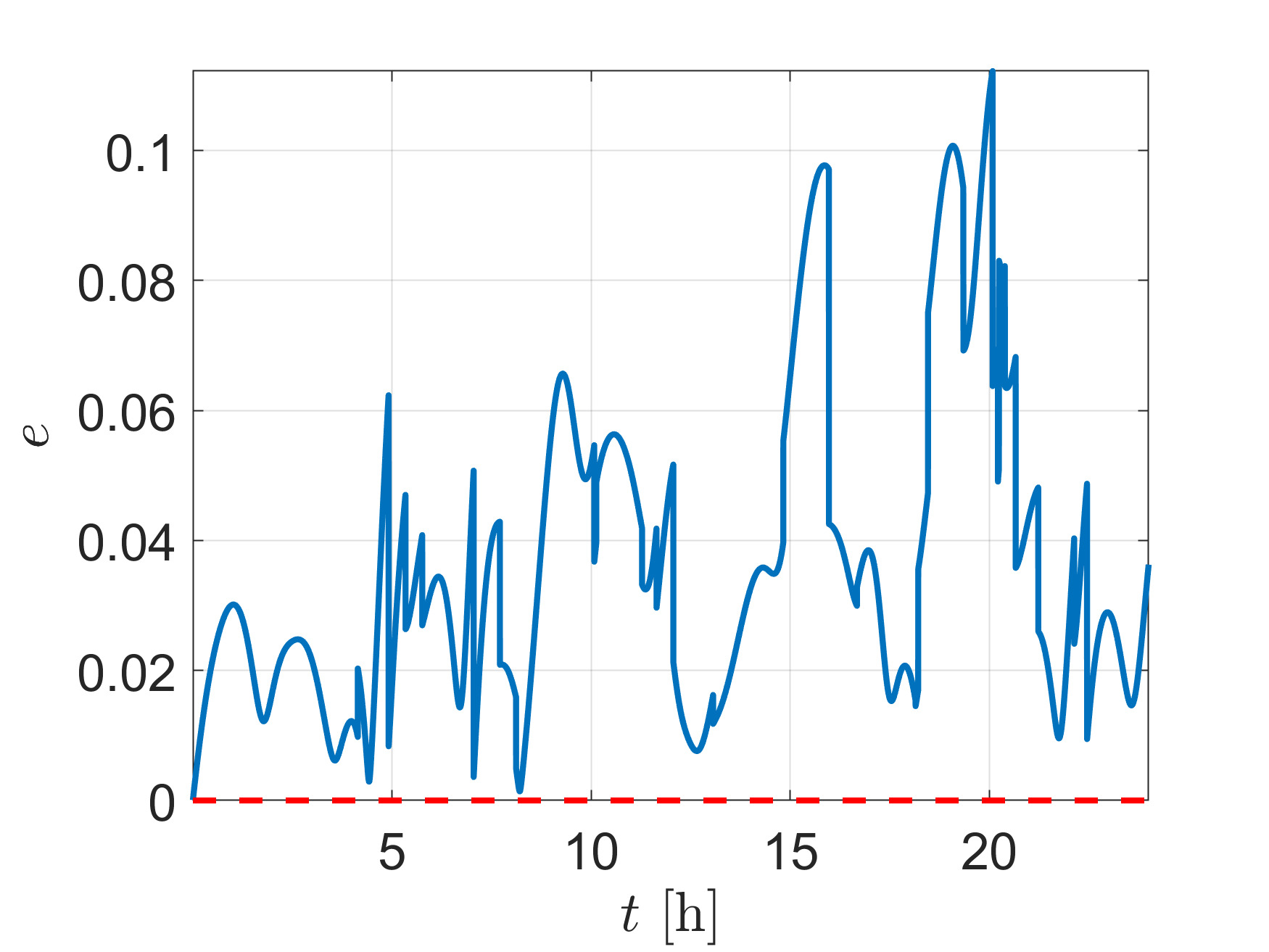}\label{Fig9d}}
\subfloat[Angular elements errors]{\includegraphics[width=.33\textwidth]{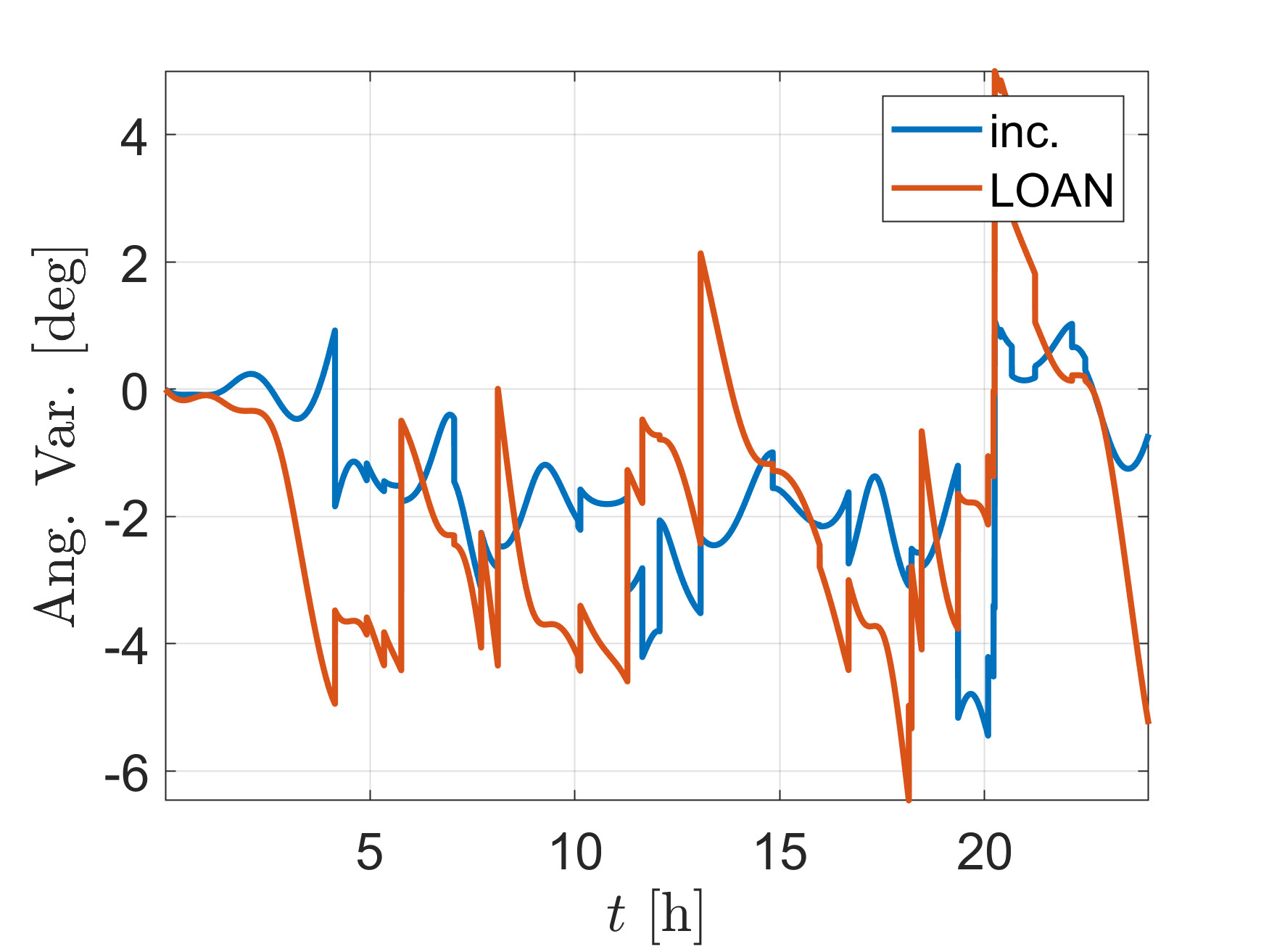}\label{Fig9e}}

\caption{Bennu looser hysteresis example, in the inertial frame.}
\label{Fig9}
\end{figure}

In a more common application, these idle-thrusters periods would be in the order of hours or days. Considering the same orbit just presented, we will loosen the parameters of the hysteresis in our proposition of control switcher, with $\vec{s}^+=\begin{bmatrix} 0.1 & 2.0 & 0.15  \end{bmatrix}^\mathbb{T}$ and $\vec{s}^-=\frac{1}{3}\vec{\Phi}$. The other control parameters are presented in Table \ref{tab:Control} as ``Bennu-loose'' example. Figure \ref{Fig9a} presents how the orbit evolves in 24 hours with this looser hysteresis. One can check in Fig. \ref{Fig9b} that the activity of the control is indeed reduced. The idle-thrusters periods are now in the order of hours. That could be even more interesting in terms of scientific outcome as close and precise scientific measurements can be made with no thrust interference. Of course, with the advantage of a safe (robust to disturbances) and autonomous operation, which would be highly demanding if made by a ground team, considering the delay in communication, orbit propagation in a high fidelity dynamical model, and other operational constraints. Moreover, this relaxed bound for the orbit significantly reduces the $\Delta V$ to 17.75 cm/s/day, which is about 0.8$\%$ of OSIRIS-REx close-proximity budget. It is worth mentioning that such long idle-thrusters periods can be best managed with a path-following law, such as the one presented here. The budget $\Delta V$ could increase in a reference tracking strategy, and a collision risk would be a pressing concern if not well handled. The orbital elements for this loose operation are shown in Figs. \ref{Fig9c} to \ref{Fig9e}.  

\begin{figure}[!htb]
\centering
\subfloat[Controlled orbit]{\includegraphics[width=.33\textwidth]{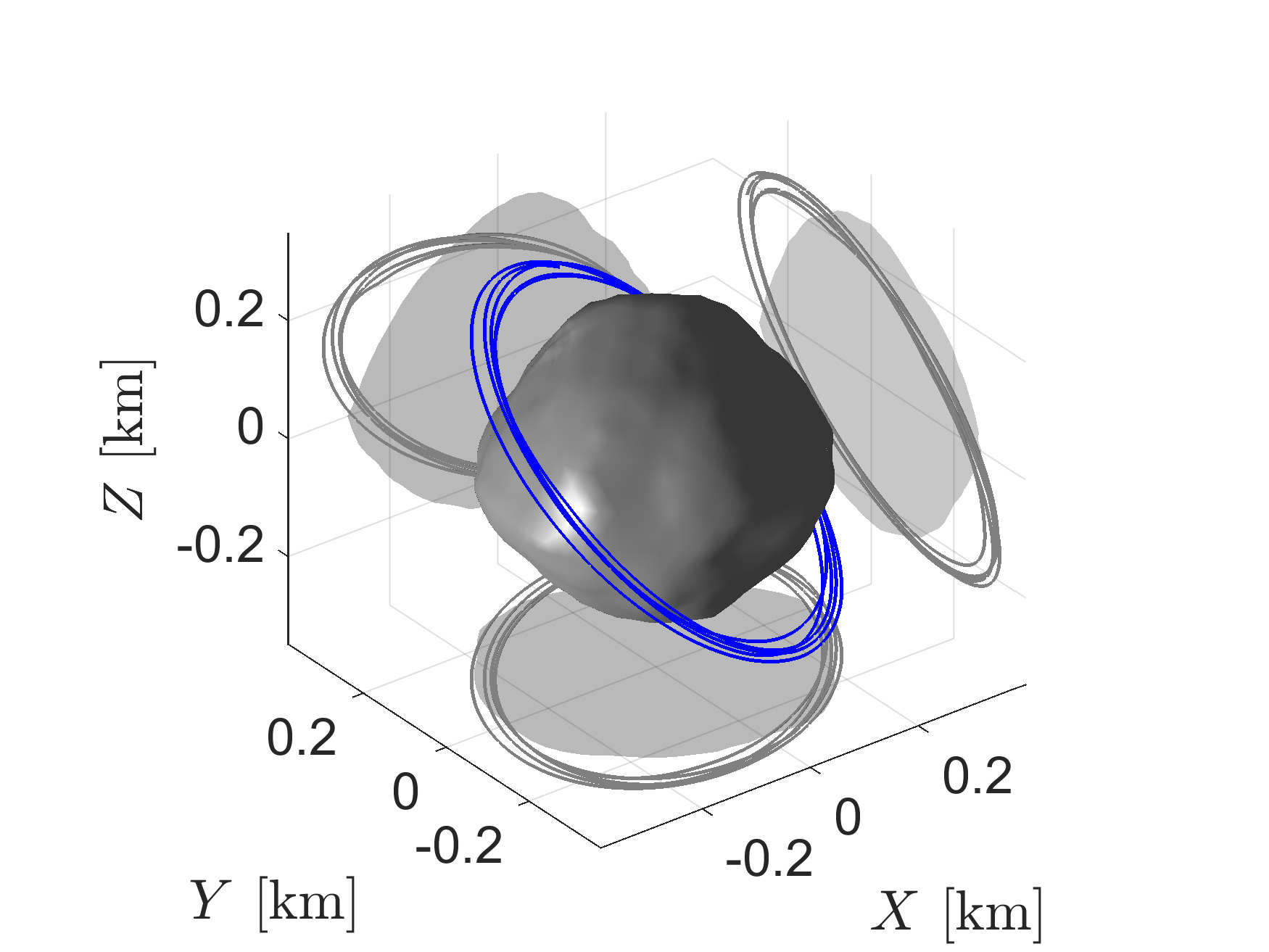}\label{Fig10a}}
\subfloat[Control commands]{\includegraphics[width=.33\textwidth]{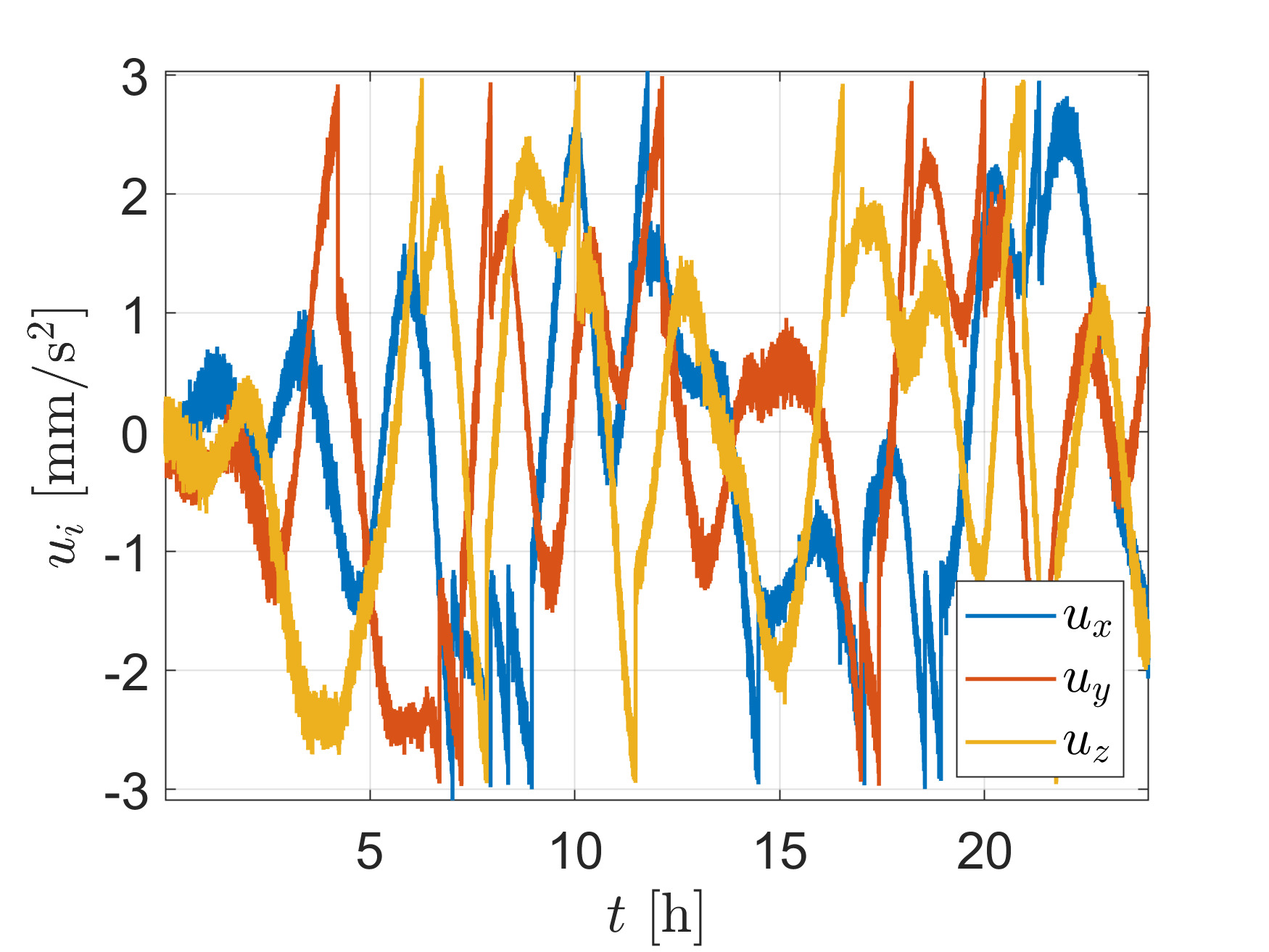}\label{Fig10b}}  
\subfloat[Modulated commands]{\includegraphics[width=.33\textwidth]{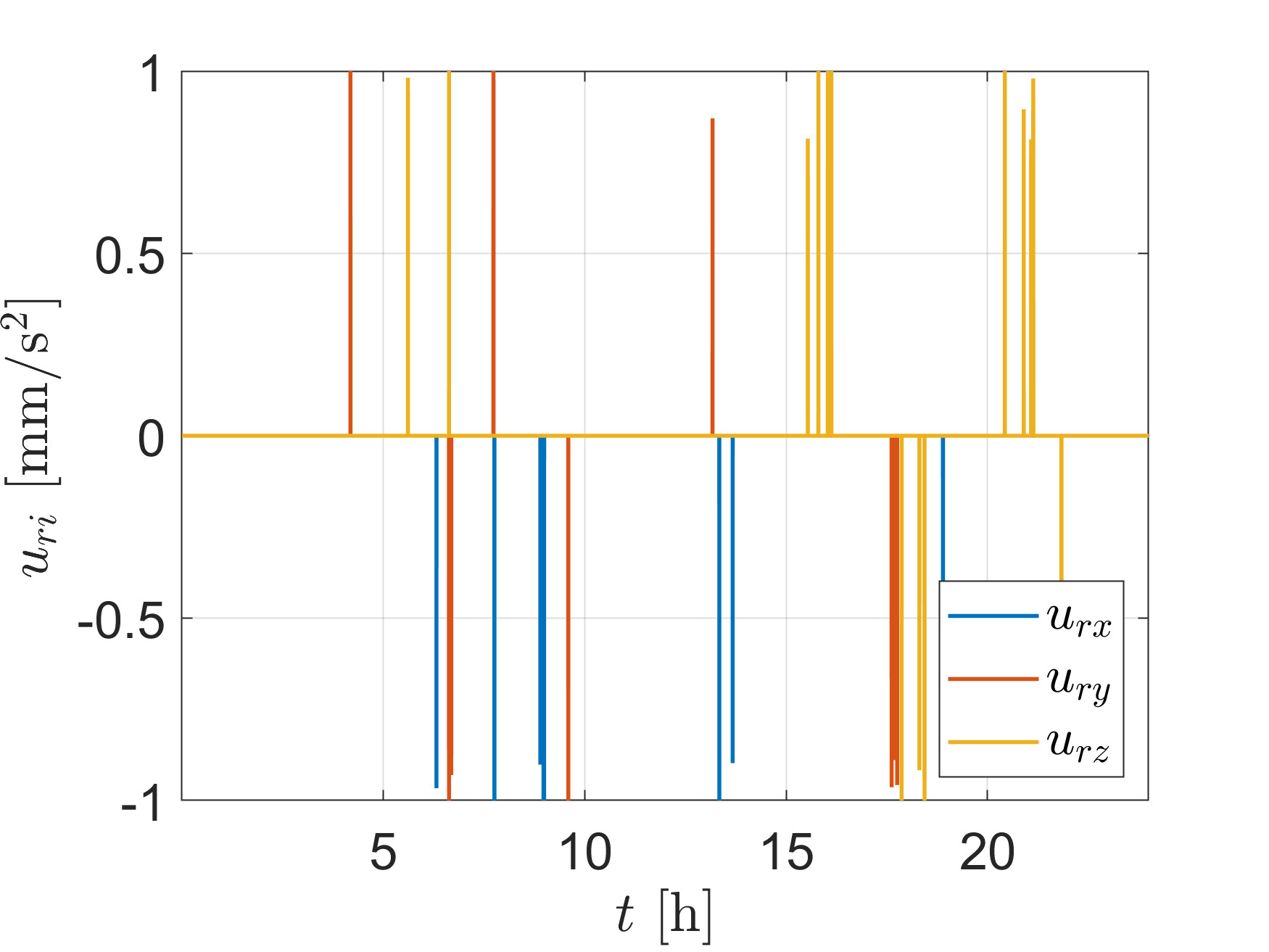}\label{Fig10c}} %\\
%\subfloat[Semi-major axis]{\includegraphics[width=.33\textwidth]{Fig10d}\label{Fig10d}}
%\subfloat[Eccentricity]{\includegraphics[width=.33\textwidth]{Fig10e}\label{Fig10e}}
%\subfloat[Angular elements errors]{\includegraphics[width=.33\textwidth]{Fig10f}\label{Fig10f}}

\caption{Bennu PWPF example, in the inertial frame.}
\label{Fig10}
\end{figure}

Finally, for the same orbit, we use now the control law in Eq. [\ref{eq:u_sat}] as the input for a pulse-width-pulse-frequency (PWPF) modulator \cite{wie2008space}. As many propulsive systems cannot deliver a continuous control input, a PWPF modulator transforms the calculated control input $u$ in a discrete signal, considering the effective thrust magnitude $u_m$ that the system can apply. For this example we consider the following PWPF parameters: $K_{LPF}=1$, $\omega_c=1$, $\delta_{on}=2.9 \times 10^{-3}$, $\delta_{off}=2.5\times 10^{-3}$, and $u_m=1$ mm/s$^2$ (details in the Appendix). The control parameters are presented in Table \ref{tab:Control} as ``Bennu-PWPF'' example. Here, it suffices to show that it works well for the proposed orbit-keeping control. The PWPF calibration is a demanding topic by itself \cite{wie2008space}. Figure \ref{Fig10a} represents the controlled orbit for this case. The calculated control command is presented in Fig. \ref{Fig10b}, and in Fig. \ref{Fig10c} the applied control is shown. Remember that we are considering an inability to execute the exact control command. That is why the modulated command is not always equal to the chosen $u_m$. The $\Delta V$ is 19.47 cm/s in the 24 hours of operation.

\begin{figure}[!htb]
\centering
\subfloat[Hohmann-like transfer]{\includegraphics[width=.33\textwidth]{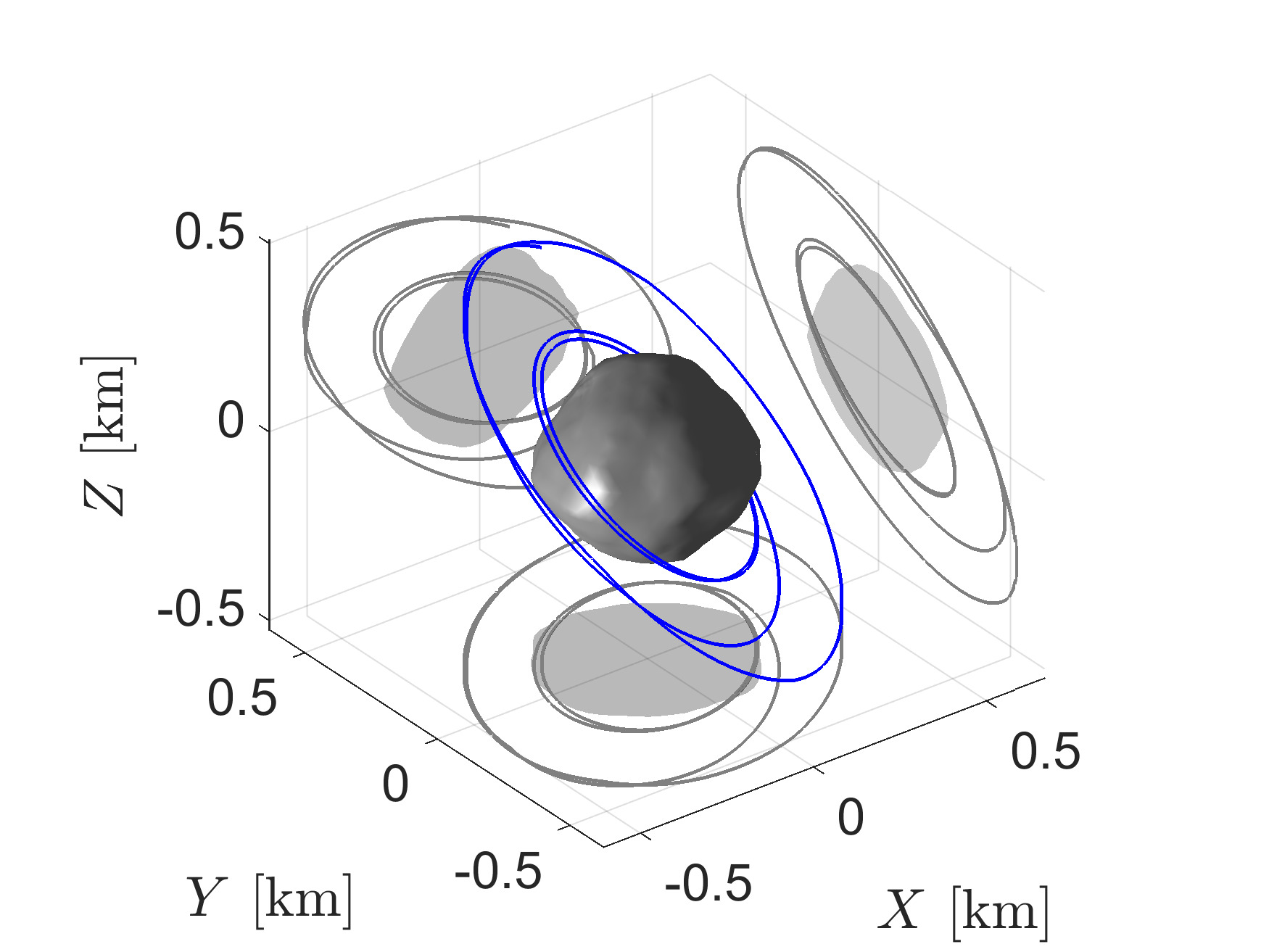}\label{Fig11a}}
\subfloat[Hyperbolic approaches]{\includegraphics[width=.33\textwidth]{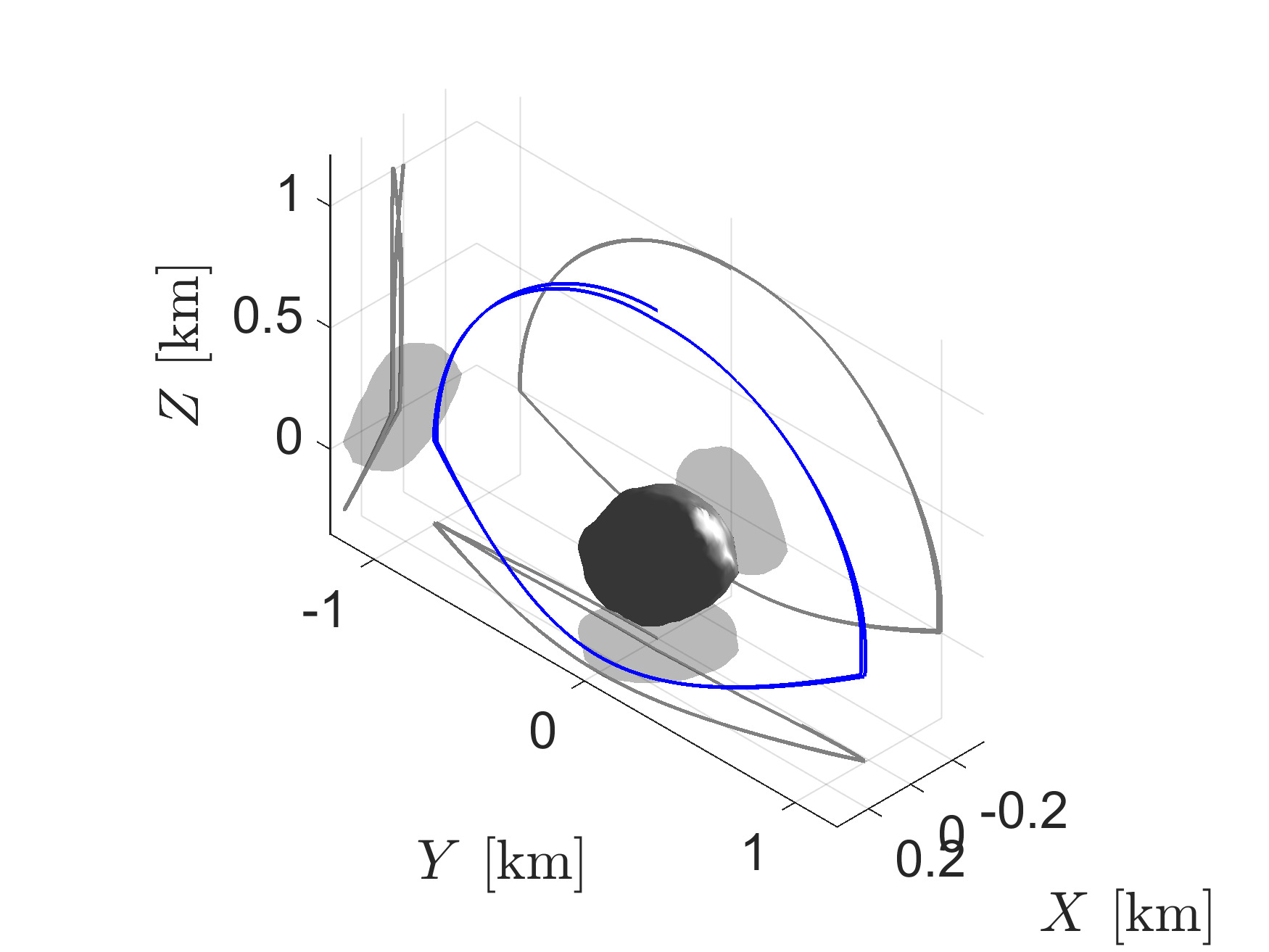}\label{Fig11b}}  
\caption{Interesting applications about Bennu, in the inertial frame.}
\label{Fig11}
\end{figure}

In Fig. \ref{Fig4}, we already showed how our orbit-keeping law could be applied to control not only natural orbits around small bodies. Now, we give some further examples of interesting applications that might exceed the orbit-keeping. We begin by presenting a Hohmann-like transfer in close-proximity to Bennu, applying the switching control in Eq. [\ref{eq:u_sat_hys}]. The control parameters are presented in Table \ref{tab:Control} as ``Bennu-Hohmann'' example. The spacecraft must stabilize a circular orbit of a radius of 600 m. Thus, after 15 hours of operation, it must depart from this circular orbit to a transfer ellipse ($a=475$ m, $e=0.2632$), culminating in a tighter circular orbit ($a=350$ m). The algorithm for that is pretty simple. After 15 hours, the desired orbital elements are changed to the ones of the transfer ellipse. Once the spacecraft is arbitrarily close to the transfer ellipse's periapsis, the orbital elements are changed to the ultimate circular orbit. Figure \ref{Fig11a} presents the simulated controlled orbit for this case, for a minimal budget $\Delta V$ of 16.23 cm/s. 

Of course, a transfer like that is not optimal. Nevertheless, even if a calculated optimal transfer reduces the $\Delta V$ to near zero, the significance of 16.23 cm/s is already minor for having a robust 30 hours operation with a transfer between two close-proximity circular orbits - which otherwise would be unstable. Moreover, in that hypothetical optimal transfer, the cost for a whole ground team operation should be added, the delay in the communication which would place operational safeness issues, and the complex modeling in this highly perturbed (and maybe not fully known) environment to obtain a safe and trustworthy optimal transfer. On the contrary, our control law would assure instantaneous response to the environment, with no need of a super complex dynamical modeling and assuring that no catastrophic output would derive from it. All of these, at the expanse of a few cm/s.

So far, we have only considered closed orbits. However, the control law is based on the two-body problem by controlling the specific angular momentum and eccentricity vectors. Thus, the proposed path-following control is applicable to obtain any conic section \cite{negri2020path}. In the last example, Figure \ref{Fig11b}, we present patching between circular and hyperbolic intersecting trajectories. The control parameters are presented in Table \ref{tab:Control} as ``Bennu-hyperbolic'' example. For positive values of $Z$, the spacecraft must stabilize a circular sun-terminator orbit with $a=1000$ m. When the value of $Z$ is negative, the spacecraft is requested to stabilize in an intersecting hyperbole with a periapsis of 400 m, $i_d=40\degree$, $\omega_d=270\degree$. The $\Delta V$ for this hypothetical operation, in a 30 hours simulation, is 57.83 cm/s. That exemplifies how the proposed orbit-keeping law can have large implications in operation close to small bodies, as it allows for a myriad of trajectories, attending operational requisites like robustness, idle-thrusters periods, low budget $\Delta V$, and safeness.

\begin{figure}[!htb]
\centering
\subfloat[Controlled orbit]{\includegraphics[width=.33\textwidth]{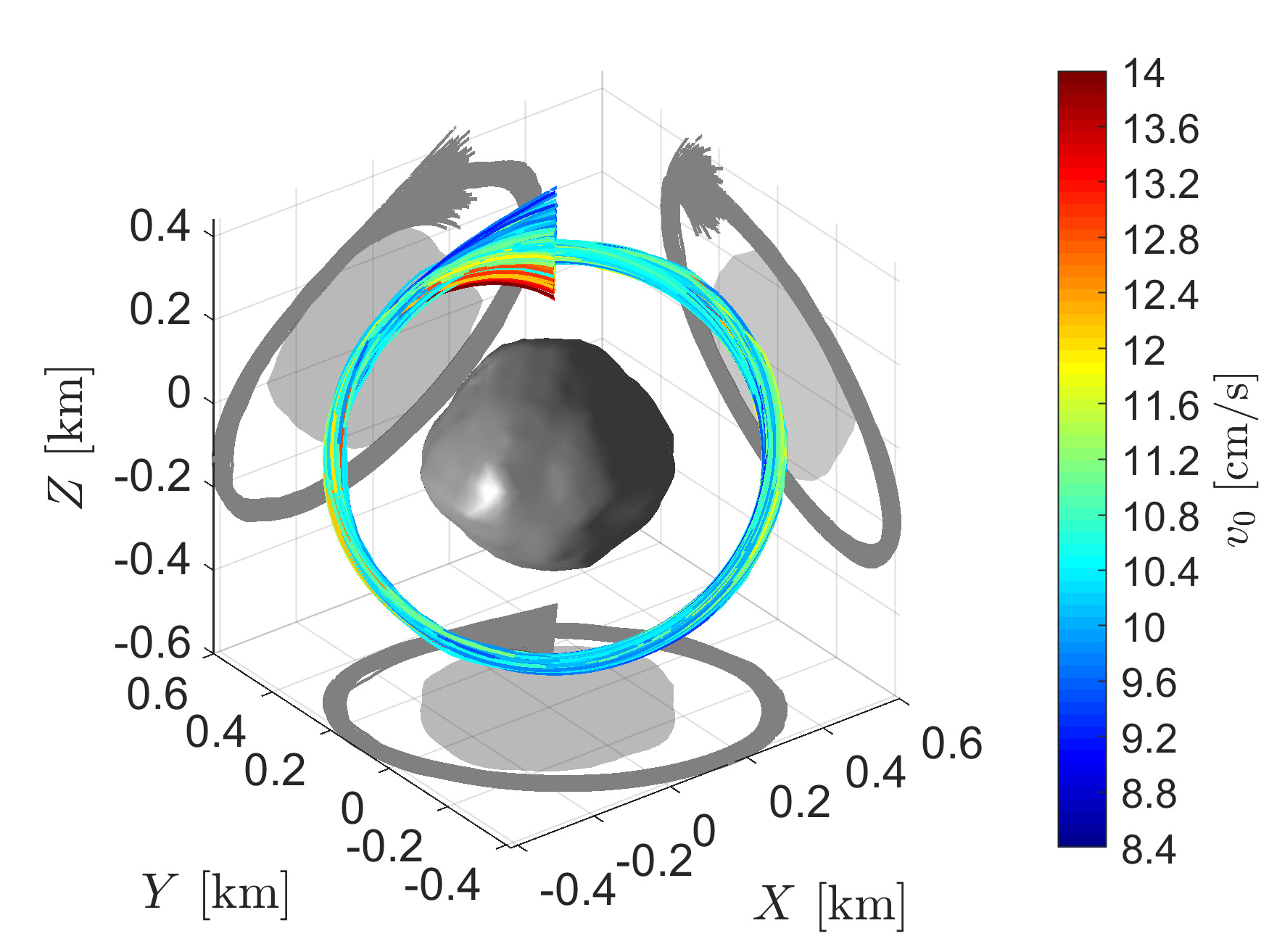}\label{Fig12a}}
\subfloat[Cummulative $\Delta V$]{\includegraphics[width=.33\textwidth]{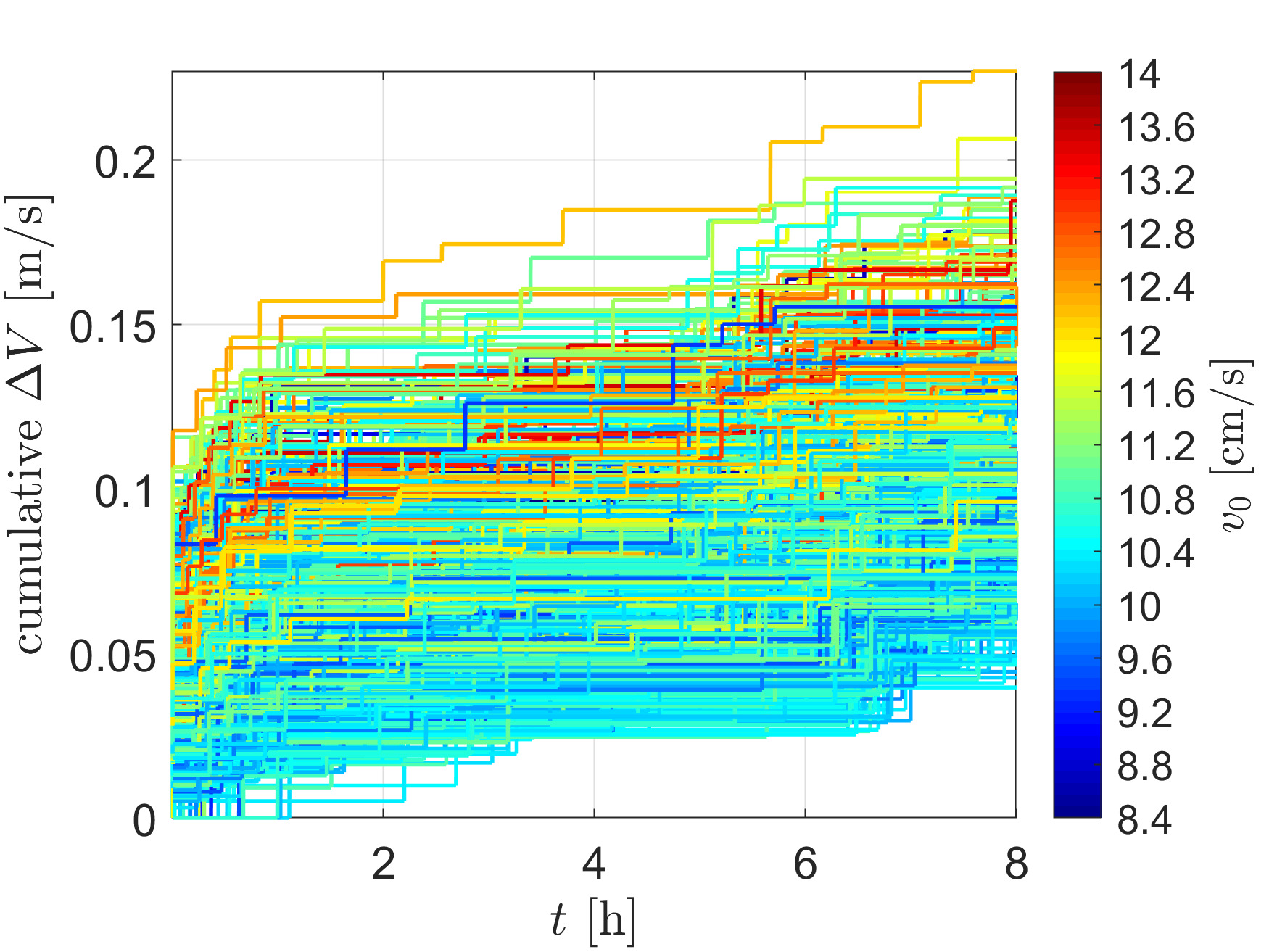}\label{Fig12b}}  
\caption{Monte Carlo simulations, orbit insertion error around Bennu, in the inertial frame.}
\label{Fig12}
\end{figure}

Finally, we consider the ability of the control to achieve the desired orbit after an orbital insertion fail. For this sake, we run a Monte Carlo simulation for 1,000 samples, using the spherical harmonics model. The desired circular orbit has $a_d=450$ m, $i_d= 45 \degree $, and $\Omega = 320 \degree$. It is assumed a 1$\sigma$ dispersion of 35 m in each orthogonal component of a plane perpendicular to the initial velocity vector expected for the desired orbit. We also assume a 1$\sigma$ dispersion of 2 cm/s in each component of what should be the initial velocity vector for the desired orbit. Note that these errors are indeed large. It goes as far as a 3$\sigma$ of 148.5 m (33$\%$ of the orbital radius) and 6 cm/s in each velocity vector component, which is about 58$\%$ of the magnitude of the desired orbital velocity (10.42 cm/s). The control parameters are presented in Table \ref{tab:Control} as ``Bennu-Monte Carlo'' example. Figure \ref{Fig12a} shows that in all of the samples the desired orbit is successfully controlled and maintained bounded due to the control switcher. Each sample trajectory is depicted in a different color, which represents the magnitude of the initial velocity vector. As one can check, it is a very extreme insertion fail condition that the control must handle. Figure \ref{Fig12b} shows the cumulative $\Delta V$ for each sample. The cumulative $\Delta V$ has a mean of 10.40 cm/s and a 3$\sigma$ of 8.60 cm/s.

\section{Conclusions}

In this work, we applied a recently derived path-following control for Keplerian orbits in the orbit-keeping problem around small bodies. The control law is straightforward and entirely analytical, being an excellent choice for a real-time operation. We considered real applications for small body missions so that the control law should be:

\begin{itemize}
    \item Proved to be robust to disturbances;
    \item Proved to be stable for any feasible operation;
    \item Adequate for any orbital geometry;
    \item Able to control only the orbit's geometry to straightforwardly and safely accommodate idle-thrusters periods.
\end{itemize}

We succeed in showing that the control law can achieve all of these conditions, allowing it to be straight applied to an actual mission scenario. We exemplified that in demanding applications with particular attention to the Bennu asteroid, which the OSIRIS-REx recently orbited. The control law is efficient in its task, even with the downgrade of replacing the discontinuous law inherent to the sliding mode control by a saturation function to avoid control chattering. The illustrative examples also suggest increasing the mission science outcome by allowing an unlimited set of orbital geometries and reducing operational costs. Although this control is applied with the small body mission application in mind, it should be remarked that it can be straight applied to any other mission with similar requirements.

\section*{Appendix - PWPF modulation}

\begin{figure}[!htb]
\centering
\subfloat[PWPF modulator]{\includegraphics[width=.5\textwidth]{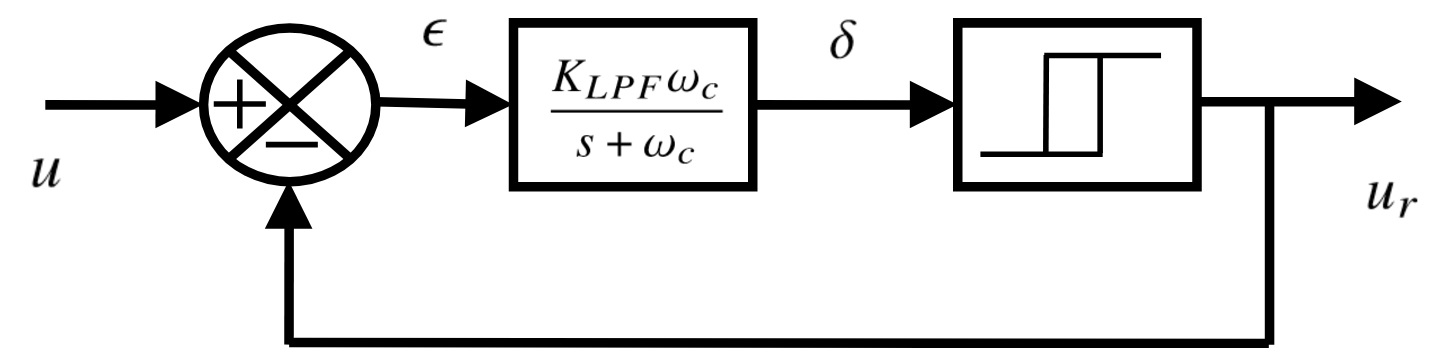}\label{Ap_Fig1a}} 
\subfloat[Schmitt trigger]{\includegraphics[width=.3\textwidth]{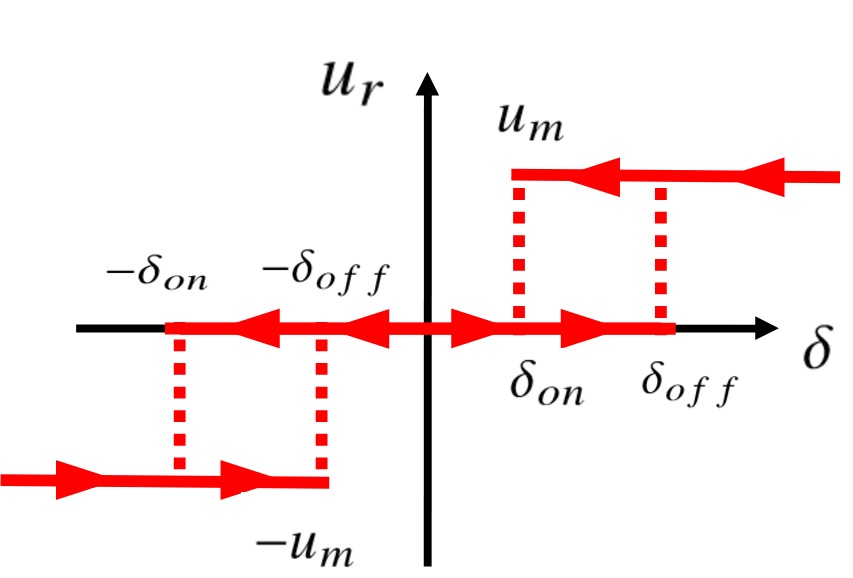}\label{Ap_Fig1b}}

\caption{Pulse-width-pulse-frequency modulator.}
\label{Ap_Fig1}
\end{figure}

Many of the thrusters currently applied in propulsion systems cannot handle a continuous control $u$. They operate in an on-off fashion by closing or opening the fuel valve and delivering, or not, a constant thrust $u_m$. In order to convert continuous control input to the discrete nature of these thrusters, a pulse-width-pulse-frequency modulation (PWPF) \cite{wie2008space,lian2013libration} is often applied. The PWPF consists of a closed-loop system with a low-pass filter of gain $K_{LPF}$ and cutoff frequency $\omega_c$, and a Schmitt trigger, which consists of a hysteresis with parameters to be adjusted $\delta_{on}$ and $\delta_{off}$, as shown in Fig. \ref{Ap_Fig1}. For more details we refer the reader to Wie \cite{wie2008space} and Lian \& Tang \cite{lian2013libration}.

\section*{Funding Sources}

The authors wish to express their appreciation for the support provided by grants $\#$ 406841/2016-0 and 301338/2016-7 from the  National Council for Scientific and Technological Development (CNPq); grants $\#$ 2017/20794-2 and 2016/24561-0 from S\~ao Paulo Research Foundation (FAPESP) and the financial support from the Coordination for the Improvement of Higher Education Personnel (CAPES).  % here.

%\section*{Acknowledgments}
%An Acknowledgments section, if used, \textbf{immediately precedes} the References. Individuals other than the authors who contributed to the underlying research may be acknowledged in this section. The use of special facilities and other resources also may be acknowledged. 

\bibliographystyle{ieeetr}
\bibliography{referencia}

\begin{thebibliography}{10}

\bibitem{guelman1994power}
M.~Guelman and D.~Harel, ``Power limited soft landing on an asteroid,'' {\em
  Journal of guidance, control, and dynamics}, vol.~17, no.~1, pp.~15--20,
  1994.

\bibitem{sawai2002control}
S.~Sawai, D.~Scheeres, and S.~Broschart, ``Control of hovering spacecraft using
  altimetry,'' {\em Journal of Guidance, Control, and Dynamics}, vol.~25,
  no.~4, pp.~786--795, 2002.

\bibitem{broschart2005control}
S.~B. Broschart and D.~J. Scheeres, ``Control of hovering spacecraft near small
  bodies: application to asteroid 25143 itokawa,'' {\em Journal of Guidance,
  Control, and Dynamics}, vol.~28, no.~2, pp.~343--354, 2005.

\bibitem{broschart2007boundedness}
S.~B. Broschart and D.~J. Scheeres, ``Boundedness of spacecraft hovering under
  dead-band control in time-invariant systems,'' {\em Journal of Guidance,
  Control, and Dynamics}, vol.~30, no.~2, pp.~601--610, 2007.

\bibitem{guelman2015closed}
M.~Guelman, ``Closed-loop control of close orbits around asteroids,'' {\em
  Journal of Guidance, Control, and Dynamics}, vol.~38, no.~5, pp.~854--860,
  2015.

\bibitem{guelman2017closed}
M.~M. Guelman, ``Closed-loop control for global coverage and equatorial
  hovering about an asteroid,'' {\em Acta Astronautica}, vol.~137,
  pp.~353--361, 2017.

\bibitem{gui2017control}
H.~Gui and A.~H.~d. Ruiter, ``Control of asteroid-hovering spacecraft with
  disturbance rejection using position-only measurements,'' {\em Journal of
  Guidance, Control, and Dynamics}, vol.~40, no.~10, pp.~2401--2416, 2017.

\bibitem{yang2017rapid}
H.~Yang, X.~Bai, and H.~Baoyin, ``Rapid generation of time-optimal trajectories
  for asteroid landing via convex optimization,'' {\em Journal of Guidance,
  Control, and Dynamics}, vol.~40, no.~3, pp.~628--641, 2017.

\bibitem{furfaro2013asteroid}
R.~Furfaro, D.~Cersosimo, and D.~R. Wibben, ``Asteroid precision landing via
  multiple sliding surfaces guidance techniques,'' {\em Journal of Guidance,
  Control, and Dynamics}, vol.~36, no.~4, pp.~1075--1092, 2013.

\bibitem{furfaro2015hovering}
R.~Furfaro, ``Hovering in asteroid dynamical environments using higher-order
  sliding control,'' {\em Journal of Guidance, Control, and Dynamics}, vol.~38,
  no.~2, pp.~263--279, 2015.

\bibitem{yang2017finite}
H.~Yang, X.~Bai, and H.~Baoyin, ``Finite-time control for asteroid hovering and
  landing via terminal sliding-mode guidance,'' {\em Acta Astronautica},
  vol.~132, pp.~78--89, 2017.

\bibitem{williams2018osiris}
B.~Williams, P.~Antreasian, E.~Carranza, C.~Jackman, J.~Leonard, D.~Nelson,
  B.~Page, D.~Stanbridge, D.~Wibben, K.~Williams, {\em et~al.}, ``Osiris-rex
  flight dynamics and navigation design,'' {\em Space Science Reviews},
  vol.~214, no.~4, p.~69, 2018.

\bibitem{scheeres2013design}
D.~Scheeres, B.~Sutter, and A.~Rosengren, ``Design, dynamics and stability of
  the osiris-rex sun-terminator orbits,'' {\em Advances in the Astronautical
  Sciences}, vol.~148, pp.~3263--3282, 2013.

\bibitem{takahashi2021autonomous}
S.~Takahashi and D.~J. Scheeres, ``Autonomous exploration of a small near-earth
  asteroid,'' {\em Journal of Guidance, Control, and Dynamics}, pp.~1--19,
  2021.

\bibitem{ohira2020autonomous}
G.~Ohira, S.~Kashioka, Y.~Takao, T.~Iyota, and Y.~Tsuda, ``Autonomous
  image-based navigation using vector code correlation algorithm for distant
  small body exploration,'' {\em Acta Astronautica}, 2020.

\bibitem{aguiar2007trajectory}
A.~P. Aguiar and J.~P. Hespanha, ``Trajectory-tracking and path-following of
  underactuated autonomous vehicles with parametric modeling uncertainty,''
  {\em IEEE transactions on automatic control}, vol.~52, no.~8, pp.~1362--1379,
  2007.

\bibitem{aguiar2008performance}
A.~P. Aguiar, J.~P. Hespanha, and P.~V. Kokotovi{\'c}, ``Performance
  limitations in reference tracking and path following for nonlinear systems,''
  {\em Automatica}, vol.~44, no.~3, pp.~598--610, 2008.

\bibitem{oguri2021robust}
K.~Oguri and J.~W. McMahon, ``Robust spacecraft guidance around small bodies
  under uncertainty: Stochastic optimal control approach,'' {\em Journal of
  Guidance, Control, and Dynamics}, pp.~1--19, 2021.

\bibitem{schaub2000spacecraft}
H.~Schaub, S.~R. Vadali, J.~L. Junkins, and K.~T. Alfriend, ``Spacecraft
  formation flying control using mean orbit elements,'' {\em Journal of the
  Astronautical Sciences}, vol.~48, no.~1, pp.~69--87, 2000.

\bibitem{garulli2011autonomous}
A.~Garulli, A.~Giannitrapani, M.~Leomanni, and F.~Scortecci, ``Autonomous
  low-earth-orbit station-keeping with electric propulsion,'' {\em Journal of
  guidance, control, and dynamics}, vol.~34, no.~6, pp.~1683--1693, 2011.

\bibitem{de2014virtual}
S.~De~Florio, S.~D’Amico, and G.~Radice, ``Virtual formation method for
  precise autonomous absolute orbit control,'' {\em Journal of Guidance,
  Control, and Dynamics}, vol.~37, no.~2, pp.~425--438, 2014.

\bibitem{negri2020path}
R.~B. Negri and A.~F. B. d.~A. Prado, ``A novel robust 3-d path following
  control for keplerian orbits,'' {\em arXiv preprint arXiv:2012.01954}, 2020.

\bibitem{scheeres2016orbital}
D.~J. Scheeres, {\em Orbital motion in strongly perturbed environments:
  applications to asteroid, comet and planetary satellite orbiters}.
\newblock Springer, 2016.

\bibitem{werner1996exterior}
R.~A. Werner and D.~J. Scheeres, ``Exterior gravitation of a polyhedron derived
  and compared with harmonic and mascon gravitation representations of asteroid
  4769 castalia,'' {\em Celestial Mechanics and Dynamical Astronomy}, vol.~65,
  no.~3, pp.~313--344, 1996.

\bibitem{dobrovolskis1996inertia}
A.~R. Dobrovolskis, ``Inertia of any polyhedron,'' {\em Icarus}, vol.~124,
  no.~2, pp.~698--704, 1996.

\bibitem{werner1997spherical}
R.~A. Werner, ``Spherical harmonic coefficients for the potential of a
  constant-density polyhedron,'' {\em Computers \& Geosciences}, vol.~23,
  no.~10, pp.~1071--1077, 1997.

\bibitem{montebruck2000satellite}
O.~Montebruck and E.~Gill, ``Satellite orbits,'' {\em Models, methods and
  applications}, 2000.

\bibitem{slotine1991applied}
J.-J.~E. Slotine, W.~Li, {\em et~al.}, {\em Applied nonlinear control},
  vol.~199.
\newblock Prentice hall Englewood Cliffs, NJ, 1991.

\bibitem{khalil2014nonlinear}
H.~K. Khalil, {\em Nonlinear control}.
\newblock Pearson Higher Ed, 2014.

\bibitem{utkin2017sliding}
V.~Utkin, J.~Guldner, and J.~Shi, {\em Sliding mode control in
  electro-mechanical systems}.
\newblock CRC press, 2017.

\bibitem{battin1999introduction}
R.~H. Battin, {\em An introduction to the mathematics and methods of
  astrodynamics, revised edition}.
\newblock American Institute of Aeronautics and Astronautics, 1999.

\bibitem{tajmar2004indium}
M.~Tajmar, A.~Genovese, and W.~Steiger, ``Indium field emission electric
  propulsion microthruster experimental characterization,'' {\em Journal of
  propulsion and power}, vol.~20, no.~2, pp.~211--218, 2004.

\bibitem{machuca2019autonomous}
P.~Machuca and J.~P. S{\'a}nchez, ``Autonomous navigation and guidance for
  cubesats to flyby near-earth asteroids,'' in {\em Proceedings of the 70th
  International Astronautical Congress (IAC), Washington DC, USA}, pp.~21--25,
  2019.

\bibitem{antreasian2019early}
P.~G. Antreasian, M.~C. Moreau, C.~D. Adam, A.~French, J.~Geeraert, K.~M.
  Getzandanner, D.~E. Highsmith, J.~M. Leonard, E.~J. Lessac-Chenen, A.~H.
  Levine, {\em et~al.}, ``Early navigation performance of the osiris-rex
  approach to bennu,'' 2019.

\bibitem{snyder2019electric}
J.~S. Snyder, D.~M. Goebel, V.~Chaplin, A.~Lopez~Ortega, I.~G. Mikellides,
  F.~Aghazadeh, I.~Johnson, T.~Kerl, and G.~Lenguito, ``Electric propulsion for
  the psyche mission,'' 2019.

\bibitem{racca2010lisa}
G.~D. Racca and P.~W. McNamara, ``The lisa pathfinder mission,'' {\em Space
  science reviews}, vol.~151, no.~1-3, pp.~159--181, 2010.

\bibitem{team2000autonomous}
A.~Team, J.~Riedel, S.~Bhaskaran, S.~Desai, D.~Han, B.~Kennedy, G.~Null,
  S.~Synnott, T.~Wang, R.~Werner, {\em et~al.}, ``Autonomous optical navigation
  (autonav) ds1 technology validation report,'' {\em Jet Propulsion Laboratory,
  California Institute of Technology}, 2000.

\bibitem{riedel2006autonav}
J.~Riedel, D.~Eldred, B.~Kennedy, D.~Kubitscheck, A.~Vaughan, R.~Werner,
  S.~Bhaskaran, and S.~Synnott, ``Autonav mark3: engineering the next
  generation of autonomous onboard navigation and guidance,'' in {\em AIAA
  Guidance, Navigation, and Control Conference and Exhibit}, p.~6708, 2006.

\bibitem{bhaskaran2012autonomous}
S.~Bhaskaran, ``Autonomous navigation for deep space missions,'' in {\em
  SpaceOps 2012}, p.~1267135, 2012.

\bibitem{wie2008space}
B.~Wie, {\em Space vehicle dynamics and control}.
\newblock American Institute of Aeronautics and Astronautics, 2008.

\bibitem{lian2013libration}
Y.~Lian and G.~Tang, ``Libration point orbit rendezvous using pwpf modulated
  terminal sliding mode control,'' {\em Advances in Space Research}, vol.~52,
  no.~12, pp.~2156--2167, 2013.

\end{thebibliography}

\end{document}